\documentclass[a4paper,11pt]{article}
\usepackage[dvips]{graphicx}
\usepackage{amsmath}
\usepackage{amssymb}
\usepackage{cite}
\usepackage{cancel}
\usepackage{fancybox}
\usepackage{ulem}
\usepackage{pifont}
\usepackage{colortbl}
\usepackage{multicol}

\usepackage{colortbl}
\usepackage{multicol}
\usepackage[all,knot]{xy}
\usepackage{longtable}
\makeatletter
 
 \@addtoreset{equation}{section}
\makeatother

\begin{document}

\begin{titlepage}

\begin{flushright}
 KYUSHU-HET-246\\
 UME-PP-022\\
\end{flushright}

\begin{center}

\vspace{1cm}
{\Large\textbf{
Pseudo-Nambu-Goldstone Dark Matter from\\ 
Non-Abelian  Gauge Symmetry
}}
\vspace{1cm}

\renewcommand{\thefootnote}{\fnsymbol{footnote}}
Hajime Otsuka${}^{1}$\footnote[2]{otsuka.hajime@phys.kyushu-u.ac.jp},
Takashi Shimomura${}^{1,2}$\footnote[3]{shimomura@cc.miyazaki-u.ac.jp},
Koji Tsumura${}^{1}$\footnote[4]{tsumura.koji@phys.kyushu-u.ac.jp},\\
Yoshiki Uchida${}^{1}$\footnote[5]{uchida.yoshiki@phys.kyushu-u.ac.jp},
and Naoki Yamatsu${}^{3}$\footnote[1]{yamatsu@phys.ntu.edu.tw}
\vspace{5mm}

\textit{
 $^1${Department of Physics, Kyushu University,\\
 744 Motooka, Nishi-ku, Fukuoka, 819-0395, Japan}\\
 $^2${Faculty of Education, University of Miyazaki,\\
1-1 Gakuen-Kibanadai-Nishi, Miyazaki 889-2192, Japan}\\
 $^3${Department of Physics, National Taiwan University,\\
Taipei, Taiwan 10617, R.O.C.}
}

\date{\today}

\abstract{
 We propose a pseudo-Nambu-Goldstone boson (pNGB)
 dark matter (DM) model based on an additional non-Abelian gauge
 symmetry $SU(2)_D$. The gauge symmetry $SU(2)_D$ is spontaneously
 broken to a global custodial symmetry $U(1)_V$ via the nonvanishing
 vacuum expectation values of $SU(2)_D$ doublet and triplet scalar
 fields.
 Due to the exact global symmetry $U(1)_V$, the lightest $U(1)_V$
 charged particle becomes stable.
 We assume that the lightest charged particle in the model is the
 charged complex pNGB, which we regard as DM.
 It avoids the strong constraints from current DM direct detection
 experiments due to the property of NGB.
 We find that
 the measured energy density of DM can be reproduced when the DM
 mass is larger than the half of the Higgs mass, where the lower limit
 generally comes from the constraint of DM invisible decay and the
 upper limit from DM direct detection experiments depends on the model
 parameters.
}

\end{center}
\end{titlepage}

\section{Introduction}

The standard model (SM) in particle physics is able to explain the
results of accelerator experiments with the exception of a few
anomalies. However, some phenomena
that cannot be explained within the scope of the SM have
emerged. One of the important issues in modern particle physics and
cosmology is the search for the identity of dark matter (DM).
The existence of dark matter (DM) has been confirmed by several
astronomical observations such as spiral galaxies
\cite{Corbelli:1999af,Sofue:2000jx}, gravitational lensing 
\cite{Massey:2010hh}, cosmic microwave background \cite{Aghanim:2018eyx},
and collision of bullet cluster \cite{Randall:2007ph}.

There are a lot of DM candidates since the nature of DM is not yet
understood. One such candidate is called the weakly interacting massive
particle (WIMP). An attractive property of WIMP DM is that it can be
generated thermally, which can be experimentally verified by introducing
a non-gravitational effect.
In order to achieve the DM's relic abundance, the mass
of WIMPs is expected to be in the range of $\mathcal{O}(10)$ GeV to
$\mathcal{O}(100)$ TeV.
Because of the non-gravitational interactions of WIMPs,
direct and indirect detections are expected. There is still no clear
signal for WIMPs, and hence direct detections yield strong constraints
on WIMP masses and interactions.

Several mechanisms in WIMP DM models are proposed to avoid the severe
constraints of the direct detection by considering e.g.,
a fermion DM with pseudo-scalar interactions
\cite{Freytsis:2010ne,Ipek:2014gua,Arcadi:2017wqi,Sanderson:2018lmj,Abe:2018emu,Abe:2019wjw}
and a pseudo-Nambu-Goldstone boson (pNGB) DM
with additional global $U(1)$ group symmetry
\cite{Barger:2010yn,Barducci:2016fue,Gross:2017dan,Balkin:2017aep,Ishiwata:2018sdi,Huitu:2018gbc,Cline:2019okt,Jiang:2019soj,Arina:2019tib,Karamitros:2019ewv,Abe:2020iph,Okada:2020zxo,Zhang:2021alu,Abe:2021jcz,Abe:2021vat,Zeng:2021moz}.
As pointed out in the original pNGB DM model \cite{Gross:2017dan}, the
DM has the property of Nambu-Goldstone (NG) mode, so the coupling of
the DM with the SM Higgs boson is proportional to its momentum. As a
result, the scattering cross sections of the DM with the SM particles
via the Higgs bosons are strongly suppressed, while  the annihilation
cross sections of the DM to the SM particles are kept.

Recently, a pNGB DM model based on gauged $G_{\rm SM}\times U(1)_{B-L}$
symmetry, which extends the softly broken $U(1)$ global symmetry
to the gauged $U(1)_{B-L}$ symmetry, was proposed
\cite{Abe:2020iph,Okada:2020zxo}, 
where $G_{\rm SM}:=SU(3)_C\times SU(2)_W\times U(1)_Y$.
The DM direct detection cross section is naturally
suppressed as the same as the original pNGB DM model. 
On the other hand, the pNGB DM decays into SM particles mediated by the
$U(1)_{B-L}$ gauge 
boson. As a result, the $U(1)_{B-L}$ symmetry breaking scale is greater
than $10^{13}$\,GeV for the DM mass $<1$\,TeV to escape the constraint
from DM stability, where the bound from gamma-ray observations is strong
as roughly the DM lifetime $\gtrsim10^{27}\,\mathrm{s}$ for two body
decays 
\cite{Baring:2015sza}. 
In addition, the $G_{\rm SM}\times U(1)_{B-L}$ pNGB DM model has been
extended to $SO(10)$ grand unified theory (GUT)
\cite{Abe:2021byq,Okada:2021qmi}. In this model, 
the vacuum expectation value (VEV) of the intermediate symmetry
breaking scale is
greater than
$10^{10}$\,GeV and the DM mass is only allowed to be slightly below half
the Higgs boson mass from the requirements of DM stability and grand
unification and also the constraints of the Higgs invisible decay and the
gamma-ray observations for DM annihilations.

The purpose of this paper is to propose a new pNGB DM model based on
non-Abelian gauge symmetry $SU(2)_D$.\footnote{
A pNGB DM model based on non-Abelian global symmetry $SU(2)_g$
and Abelian gauge symmetry $U(1)_X$ has been proposed in
Ref.~\cite{Abe:2022mlc}. This model can be regarded as a low-energy
effective model that is realized in a special parameter region in our
model.}
Unlike the $G_{\rm SM}\times U(1)_{B-L}$ and $SO(10)$ pNGB DM model,
we will confirm that the DM is stabilized due to the residual $U(1)$
symmetry of the $SU(2)$ custodial symmetry \cite{Sikivie:1980hm}.\footnote{
A DM model using custodial symmetry emerging from non-Abelian gauge
symmetry $SU(2)_D$ for DM stability has been proposed in
Ref.~\cite{Ma:2022bcj}, although it is not a pNGB DM model.
}
We will show that in our pNGB DM model the VEV of $SU(2)_D$ breaking
scale can be allowed to be roughly ${\cal O}(1)$\,TeV without
introducing very high energy scale.

The paper is organized as follows.  
In Sec.~\ref{Sec:Model}, we introduce an $SU(2)$ pNGB DM model.
In Sec.~\ref{Sec:Vacuum-structure},
we analyze vacuum structures and symmetry breaking patterns of the
model.
In Sec.~\ref{Sec:Potential},
we analyze the scalar potential of the system.
In Sec.~\ref{Sec:Mass-spectrum},
we investigate the mass spectra of scalar fields in this model.
In Sec.~\ref{Sec:Scattering},
we examine the constraints from direct detection experiments and the
thermal relic abundance of DM for our DM candidate.
Section~\ref{Sec:Summary} is devoted to
summary and discussions.
In Appendix~\ref{Sec:DM-quark-scattering}, we show the detailed
calculation of DM-quark scattering amplitude.

\section{The model}
\label{Sec:Model}

The model consists of
the SM gauge fields, 
an $SU(2)_D$ gauge field ${W}_\mu^{\prime a}$ $(a=1,2,3)$,
a complex scalar field in ${\bf 2}$ of $SU(2)_D$
$\Phi$,
and
a real scalar field in ${\bf 3}$ of $SU(2)_D$
$\Delta$.
The matter content in the non-Abelian pNGB DM model is summarized in
Table~\ref{Tab:Matter_content}.

\begin{table}[thb]
\begin{center}
\begin{tabular}{|c||c|c|c|c|c||c|c|c||c|c|c|c|}\hline
 \rowcolor[gray]{0.8}
 & $Q$ & $u^{c}$ & $d^{c}$
 & $L$ & $e^{c}$
 & $H$ 
 & $\Phi$
 & $\Delta$
 & $G_{\mu\nu}$
 & $W_{\mu\nu}$
 & $B_{\mu\nu}$
 & $W_{\mu\nu}'$
 \\\hline\hline
 $SU(3)_{C}$
 & ${\bf 3}$ & ${\bf \overline{3}}$ & ${\bf \overline{3}}$
 & ${\bf 1}$ & ${\bf 1}$ 
 & ${\bf 1}$ & ${\bf 1}$ & ${\bf 1}$ 
 & ${\bf 8}$ & ${\bf 1}$ & ${\bf 1}$ & ${\bf 1}$ 
 \\ \hline
 $SU(2)_{W}$
 & ${\bf 2}$ & ${\bf 1}$ & ${\bf 1}$
 & ${\bf 2}$ & ${\bf 1}$ 
 & ${\bf 2}$ & ${\bf 1}$ & ${\bf 1}$ 
 & ${\bf 1}$ & ${\bf 3}$ & ${\bf 1}$ & ${\bf 1}$ 
 \\ \hline
 $U(1)_{Y}$
 & $+1/6$ & $-2/3$ & $+1/3$
 & $-1/2$ & $+1$  
 & $+1/2$  & $0$  & $0$
 & $0$ & $0$ & $0$ & $0$				     
 \\\hline
 $SU(2)_{D}$
 & ${\bf 1}$ & ${\bf 1}$ & ${\bf 1}$
 & ${\bf 1}$ & ${\bf 1}$
 & ${\bf 1}$ & ${\bf 2}$ & ${\bf 3}$
 & ${\bf 1}$ & ${\bf 1}$ & ${\bf 1}$ & ${\bf 3}$ 
 \\ \hline
\end{tabular}
 \caption{\small The field content in the pNGB DM model 
 is shown in the $G_{\rm SM}\times SU(2)_D$ basis,
 where the fermions belong to $(1/2,0)$ under $SL(2,\mathbb{C})$.
 }
\label{Tab:Matter_content}
\end{center}  
\end{table}

The Lagrangian is given by
\begin{align}
 {\cal L}&=
  -\frac{1}{2}\mbox{tr}
 \left[{G}_{\mu\nu}{G}^{\mu\nu}\right]
  -\frac{1}{2}\mbox{tr}
 \left[{W}_{\mu\nu}{W}^{\mu\nu}\right]
  -\frac{1}{4}
 {B}_{\mu\nu}{B}^{\mu\nu}
  -\frac{1}{2}\mbox{tr}
 \left[{W}_{\mu\nu}^{\prime}{W}^{\prime\mu\nu}\right]
 \nonumber\\
 &\hspace{1em}
 +\left({D}_\mu H\right)^\dag
 \left({D}^\mu H\right)
 +\left({D}_\mu{\Phi}\right)^\dag
 \left({D}^\mu{\Phi}\right)
 +\frac{1}{2}
 \mbox{tr}\left[
 \left({D}_\mu\Delta\right)
 \left({D}^\mu\Delta\right)\right]
 \nonumber\\
 &\hspace{1em}
 -{\cal V}\left(H,\Phi,\Delta\right)
 \nonumber\\
 &\hspace{1em}
 +\overline{Q}i\cancel{{D}}Q
 +\overline{u^c}i\cancel{{D}}u^c
 +\overline{d^c}i\cancel{{D}}d^c
 +\overline{L}i\cancel{{D}}L
 +\overline{e^c}i\cancel{{D}}e^c
 \nonumber\\
 &\hspace{1em}
 -\left(y_uQu^c H+y_dQd^c H^\dag+y_eLe^cH^\dag
 +\mbox{h.c.}\right),
\label{Eq:Lagrangian}
\end{align}
where ${D}_\mu=\partial_\mu+ig_s{G}_\mu+ig_2W_\mu+ig_1B_\mu+ig_2'W_\mu'$;
${F}_{\mu\nu}=\partial_\mu{F}_\nu-\partial_\nu{F}_\mu+i{g}[{F}_\mu,{F}_\nu]$,
where $F=G,W,B,W'$ and $g_s$, $g_2$, $g_1$, $g_1'$ are 
gauge fields and gauge coupling constants of $SU(3)_C$, $SU(2)_W$,
$U(1)_Y$, $SU(2)_D$, respectively.
The scalar potential
${\cal V}\left(H,\Phi,\Delta\right)$
contains quadratic, cubic, and quartic coupling terms,
\begin{align}
 {\cal V}(H,\Phi,\Delta)&=
 -\mu_H^2H^\dag H
 -\mu_{\Phi}^2\Phi^\dag\Phi
 -\frac{1}{2}\mu_{\Delta}^2\mbox{Tr}\left(\Delta^2\right)
 \nonumber\\
 &\hspace{1em}
 +
 \sqrt{2}
 \left(
  (\kappa_1+i\kappa_2)
 \tilde{\Phi}^\dag\Delta\Phi
 + (\kappa_1-i\kappa_2)
 {\Phi}^\dag\Delta\tilde{\Phi}
 \right)
 +2\sqrt{2}\kappa_3\Phi^\dag\Delta\Phi
 \nonumber\\
 &\hspace{1em}
 +\lambda_{H}\left(H^\dag H\right)^2
 +\lambda_{\Phi}\left(\Phi^\dag \Phi\right)^2
 +\frac{1}{4}\lambda_{\Delta}\mbox{Tr}\left(\Delta^2\right)^2
 \nonumber\\
 &\hspace{1em}
 +\lambda_{H\Phi} \left(H^\dag H\right) \left(\Phi^\dag \Phi\right)
 +\lambda_{H\Delta}\left(H^\dag H\right)
 \mbox{Tr}\left(\Delta^2\right)
 +\lambda_{\Phi\Delta}
 \left(\Phi^\dag \Phi\right)
  \mbox{Tr}\left(\Delta^2\right),
\label{Eq:Potential-scalar}
\end{align}
where $\tilde{\Phi}(x)=i\sigma_2\Phi(x)^*$;
$\mu_H^{2}$, $\mu_\Phi^{2}$, and $\mu_\Delta^{2}$
are real parameters with dimension 2, 
$\kappa_a (a=1,2,3)$ are real parameters with dimension 1,
and $\lambda_{H}$, $\lambda_{\Phi}$, $\lambda_{\Delta}$
$\lambda_{H\Phi}$, $\lambda_{H\Delta}$, and $\lambda_{\Phi\Delta}$
are dimensionless real parameters.
We use the following notation:
\begin{align}
 \Delta&=
 \frac{1}{\sqrt{2}}\left(
 \begin{array}{cc}
  \eta_3&\eta_1-i\eta_2 \\
  \eta_1+i\eta_2 &-\eta_3\\
 \end{array}
 \right),
\nonumber\\
 \Phi&=
 \frac{1}{\sqrt{2}}\left(
 \begin{array}{c}
  \phi_1+i\phi_2\\
  \phi_3+i\phi_4\\
 \end{array}
 \right).
\end{align}
Under the $SU(2)_D$ transformation, $\Phi(x)$ and $\Delta(x)$
behave as 
\begin{align}
 &\Phi(x)\to U(x) \Phi(x),\ \
 \Delta(x)\to U(x)\Delta(x) U(x)^{\dag},\
\end{align}
where $U(x)$ is the $SU(2)_D$ unitary transformation
$U(x)=\mbox{exp}\left[i\theta_a(x)\frac{\sigma_a}{2}\right]$;
$\theta_a(x) (a=1,2,3)$ are the parameters of the $SU(2)_D$ gauge
transformation and $\sigma_a$ stand for the Pauli matrices.
Note that it is easy to check invariant terms under
$G_{\rm SM}\times SU(2)_D$ by using GroupMath \cite{Fonseca:2020vke} and 
Sym2Int\cite{Fonseca:2017lem,Fonseca:2019yya}.
(For Lie groups, see e.g., Ref.~\cite{Yamatsu:2015gut}.)

We will analyze the relations between vacuum structures and symmetry
breaking patterns in the next section.
In the model, the invariant terms that contain only the scalar field
$\Phi(x)$ in ${\bf 2}$ of $SU(2)_D^{\rm local}$ are invariant under 
a larger global symmetry
$SU(2)_{\Phi L}^{\rm global}\times SU(2)_{\Phi R}^{\rm global}$.
To check this extended global symmetry, it is convenient to introduce
a bi-doublet or $2\times 2$ matrix notation for $\Phi(x)$ as
\begin{align}
 {\Sigma}(x):=
 \left(
 \begin{array}{cc}
 \tilde{\Phi}(x)&\Phi(x) \\
 \end{array}
 \right).
\end{align}
This notation is convenient to understand so-called $SU(2)$ custodial
symmetry \cite{Sikivie:1980hm}.
We will find that the stability of the DM is realized by a ``
$U(1)_{V}^{\rm global}$ custodial symmetry,'' which is a $U(1)$
subgroup of the
$SU(2)_{\Phi L}^{\rm global}\times SU(2)_{\Phi R}^{\rm global}$ diagonal
subgroup $SU(2)_{V}^{\rm global}$.

By using the complex scalar field in ${\bf 2}$ of $SU(2)_D$ $\Sigma(x)$
instead of $\Phi(x)$, the scalar potential in
Eq.~(\ref{Eq:Potential-scalar}) can be written as  
\begin{align}
 {\cal V}(H,{\Sigma},\Delta)
 &=
 -\mu_H^2H^\dag H
 -\frac{\mu_\Phi^2}{2}
 \mbox{Tr}\left({\Sigma}^\dag{\Sigma}\right)
 -\frac{1}{2}\mu_{\Delta}^2\mbox{Tr}\left(\Delta^2\right)
 \nonumber\\
 &\hspace{1em}
 -
 \sqrt{2}
 \kappa_1
 \mbox{Tr}
 \left(
 \sigma_1
 {\Sigma}^\dag \Delta{\Sigma}
 \right)
 -
 \sqrt{2}
 \kappa_2
 \mbox{Tr}
 \left(
 \sigma_2
 {\Sigma}^\dag \Delta{\Sigma}
 \right)
 -\sqrt{2}\kappa_3
 \mbox{Tr}
 \left(
 \sigma_3
 {\Sigma}^\dag \Delta{\Sigma}
 \right)
 \nonumber\\
 &\hspace{1em}
 +\lambda_{H}\left(H^\dag H\right)^2
 +\frac{\lambda_\Phi}{4}
 \left(\mbox{Tr}\left({\Sigma}^\dag{\Sigma}\right)\right)^2
 +\frac{1}{4}\lambda_{\Delta}\mbox{Tr}\left(\Delta^2\right)^2
 \nonumber\\
 &\hspace{1em}
 +\frac{1}{2}\lambda_{H\Phi}
 \left(H^\dag H\right)
 \mbox{Tr}\left({\Sigma}^\dag{\Sigma}\right)
 +\frac{1}{2}\lambda_{H\Delta}\left(H^\dag H\right)
 \mbox{Tr}\left(\Delta^2\right)
 +\frac{1}{2}\lambda_{\Phi\Delta}
 \mbox{Tr}\left({\Sigma}^\dag{\Sigma}\right)
  \mbox{Tr}\left(\Delta^2\right),
\label{Eq:Potential-scalar-check}
\end{align}
where we used a relation $(-i\sigma_2)\Sigma^*(i\sigma_2)=\Sigma$.

\begin{figure}[tbh]
\begin{center}
\includegraphics[bb=0 0 667 75,height=1.5cm]{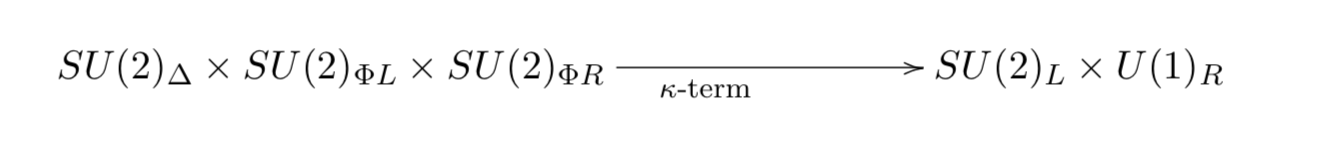}
\end{center}
 \caption{\small The global symmetry
 $SU(2)_{\Delta}^{\rm global}\times
 SU(2)_{\Phi L}^{\rm global}\times SU(2)_{\Phi R}^{\rm global}$ breaking
 pattern is shown. 
 The $\kappa$ term stands for the soft symmetry breaking term.
 In the figure, the superscript, global, is omitted. 
}
\label{Figure:Global-Symmetry-Breaking-Soft-Breaking-Terms}
\end{figure}

We verify what kind of global symmetry exists in
the potential given in Eq.~(\ref{Eq:Potential-scalar-check}).
First, the potential ${\cal V}(0,{\Sigma},0)$ is invariant under
$SU(2)_{\Phi L}^{\rm global}\times SU(2)_{\Phi R}^{\rm global}$.
$\Sigma$ is the bi-doublet representation under 
$SU(2)_{\Phi L}^{\rm global}\times SU(2)_{\Phi R}^{\rm global}$:
\begin{align}
 {\Sigma}\to
 U_{\Phi L}{\Sigma}U_{\Phi R}^\dag,\ \ \
 U_{\Phi L}=e^{i\theta_{\Phi L}^a\sigma_a},\ \
 U_{\Phi R}=e^{i\theta_{\Phi R}^a\sigma_a},
\end{align}
where $\theta_{\Phi L}^a$ and $\theta_{\Phi R}^a$ are parameters of
$SU(2)_{\Phi L}^{\rm global}$ and $SU(2)_{\Phi R}^{\rm global}$
transformations, respectively.
Second, the potential ${\cal V}(0,0,\Delta)$ is invariant under
$SU(2)_{\Delta}^{\rm global}$ transformation.
$\Delta$ is the adjoint representation under
 $SU(2)_{\Delta}^{\rm global}$:
\begin{align}
 \Delta \to U_{\Delta}\Delta U_{\Delta}^\dag,\ \
 U_{\Delta}=e^{i\theta_{\Delta}^a\sigma_a},
\end{align}
where $\theta_{\Delta}^a$ is a parameter of
$SU(2)_{\Delta}^{\rm global}$, and
the global transformation corresponds to a global subgroup
transformation of the gauge group $SU(2)_D^{\rm local}$ transformation.
Third, under
$SU(2)_{\Delta}^{\rm global}\times SU(2)_{\Phi L}^{\rm global}\times
SU(2)_{\Phi R}^{\rm global}$,
each $\kappa_{a}$ term transforms as
\begin{align}
& \mbox{Tr}
 \left(
 \sigma_a
 {\Sigma}^\dag \Delta{\Sigma}
 \right)
\to
 \mbox{Tr}
 \left(
 U_{\Phi R}^\dag
 \sigma_a
 U_{\Phi R}
 {\Sigma}^\dag
 U_{\Phi L}^\dag
 U_\Delta
 \Delta
 U_\Delta^\dag
 U_{\Phi L}
 {\Sigma}
 \right).
\end{align}
This term is invariant under 
$SU(2)_{L}^{\rm global}\times U(1)_{\Phi R{a}}^{\rm global}$,
where $SU(2)_{L}^{\rm global}$ represents
$SU(2)_\Delta=SU(2)_{\Phi L}$, and 
$U(1)_{\Phi R{a}}^{\rm global}$ corresponds to the $\sigma_a$
direction of $SU(2)_{\Phi R}^{\rm global}$.
The combination of the $\kappa_1$, $\kappa_2$, and $\kappa_3$ terms is
also invariant under
$SU(2)_{L}^{\rm global}\times U(1)_{\Phi R}^{\rm global}$: 
\begin{align}
& \mbox{Tr}
 \left(
 (\kappa_1\sigma_1+\kappa_2\sigma_2+\kappa_3\sigma_3)
 {\Sigma}^\dag \Delta{\Sigma}
 \right)
\to
 \mbox{Tr}
 \left(
 U_{\Phi R}^\dag
 (\kappa_1\sigma_1+\kappa_2\sigma_2+\kappa_3\sigma_3)
 U_{\Phi R}
 {\Sigma}^\dag\Delta{\Sigma}
 \right),
\end{align}
where the $U(1)_{\Phi R}^{\rm global}$ transformation is associated with
the $(\kappa_1\sigma_1+\kappa_2\sigma_2+\kappa_3\sigma_3)$ direction in
$SU(2)_{\Phi R}^{\rm global}$.
Therefore, the potential is invariant under 
$SU(2)_{L}^{\rm global}\times U(1)_{\Phi R}^{\rm global}$.
Without losing generality, we can choose the $\kappa_a\sigma_a$ direction
associated with the remaining $U(1)_{\Phi R}^{\rm global}$
symmetry.  
Therefore, in the following we will take the $\kappa_3\sigma_3$
direction and denote $\kappa_3$ as $\kappa$ and
$U(1)_{\Phi R3}^{\rm global}$ as $U(1)_{R}^{\rm global}$. In other words,
we remove $\kappa_1$ and $\kappa_2$ by using the
$SU(2)_{\Phi R}^{\rm global}$ transformation.
Therefore, the potential in Eq.~(\ref{Eq:Potential-scalar-check}) is
invariant under $SU(2)_L^{\rm global}\times U(1)_R^{\rm global}$.
The global symmetry
$SU(2)_{\Delta}^{\rm global}\times
SU(2)_{\Phi L}^{\rm global}\times SU(2)_{\Phi R}^{\rm global}$
breaking pattern associated with the explicit breaking terms
is shown in 
Figure~\ref{Figure:Global-Symmetry-Breaking-Soft-Breaking-Terms}. 

\section{Vacuum structure}
\label{Sec:Vacuum-structure}

\begin{figure}[tbh]
\begin{center}
\includegraphics[bb=0 0 387 132,height=2.5cm]{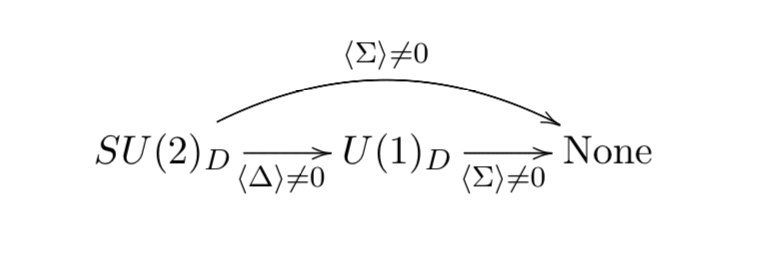}
\end{center}
 \caption{\small The gauge symmetry
 $SU(2)_{D}^{\rm local}$ breaking patterns
 are shown. $\langle{\Sigma}\rangle\not=0$ and
 $\langle\Delta\rangle\not=0$ 
 represent  the spontaneous symmetry breaking (SSB) by the VEV of ${\Sigma}$ and $\Delta$ in ${\bf
 2}$ and ${\bf 3}$ of $SU(2)_D^{\rm local}$.
 $U(1)_D$ is a local subgroup of $SU(2)_D^{\rm local}$.
 In the figure, the superscript, local, is omitted. 
}
\label{Figure:Gauge-Symmetry-Breaking}
\end{figure}

We consider vacuum structures of ${\Sigma}(x)$ and $\Delta(x)$. The
system we are currently considering has $SU(2)_{D}^{\rm local}$
(or $SU(2)_{L}^{\rm global}$) and $U(1)_{R}^{\rm global}$ symmetry. 
By using a total of four degrees of freedom of $SU(2)_{D}^{\rm local}$
gauge and $U(1)_{R}^{\rm global}$ transformations,
without loss of generality, we take the VEVs of ${\Sigma}$ and $\Delta$ as
$\langle\Delta\rangle=(v_{\eta_1}\sigma_1+v_{\eta_3}\sigma_3)/\sqrt{2}$
and 
$\langle\Sigma\rangle=v_\Phi I/\sqrt{2}$, i.e.,
\begin{align}
 \langle\Delta\rangle&=\frac{1}{\sqrt{2}}\left(
 \begin{array}{cc}
  v_{\eta_3}&v_{\eta_1}\\
  v_{\eta_1}&-v_{\eta_3}\\
 \end{array}
 \right),\ \ \
 \langle{\Sigma}\rangle=
 \frac{1}{\sqrt{2}}
 \left(
 \begin{array}{cc}
  v_\Phi&0\\
  0&v_\Phi\\
 \end{array}
 \right),
\label{Eq:VEV-full}
\end{align}
where we remove the VEVs of $\eta_2$, $\phi_1$, $\phi_2$, and $\phi_4$.
The gauge symmetry
$SU(2)_{D}^{\rm local}$ breaking patterns are shown
in Figure~\ref{Figure:Gauge-Symmetry-Breaking}.
As is well-known, the nonvanishing VEV of a complex scalar field in
${\bf 2}$ of $SU(2)$ breaks $SU(2)$ symmetry completely, so
a total of three Nambu-Goldstone (NG) or pseudo NG (pNG) modes appear.
More specifically, when $SU(2)$ global symmetry is exact, three NG modes
appear; when $SU(2)$ global symmetry is softly broken to $U(1)$ global
symmetry by explicit breaking terms, one NG and two pNG modes appear; 
when $SU(2)$ global symmetry is completely softly broken by explicit
breaking terms, three pNG modes appear.
The nonvanishing VEV of a real scalar field in ${\bf 3}$ of $SU(2)$
breaks $SU(2)$ symmetry to $U(1)$ symmetry, so a total of two NG or pNG
modes appear.

Furthermore, when $\kappa=0$, the system has 
$SU(2)_{\Delta}^{\rm global}\times
 SU(2)_{\Phi L}^{\rm global}\times
 SU(2)_{\Phi R}^{\rm global}$ symmetry.
 By using the degrees of freedom of $SU(2)_{\Delta}^{\rm global}$,
 $SU(2)_{\Phi L}^{\rm global}$, and $SU(2)_{\Phi R}^{\rm global}$
 transformations, we can take the VEVs of $\Sigma$ and $\Delta$ as 
\begin{align}
 \langle\Delta\rangle&=\frac{1}{\sqrt{2}}\left(
 \begin{array}{cc}
  v_{\eta_3}&0\\
  0&-v_{\eta_3}\\
 \end{array}
 \right),\ \ \
 \langle{\Sigma}\rangle=
 \frac{1}{\sqrt{2}}
 \left(
 \begin{array}{cc}
  v_\Phi&0\\
  0&v_\Phi\\
 \end{array}
 \right),
\label{Eq:VEV-reduce}
\end{align}

Here we check what kind of symmetry is preserved by the VEVs of
${\Sigma}$ and $\Delta$ in Eqs.~(\ref{Eq:VEV-full}) and
(\ref{Eq:VEV-reduce}).
First, we consider the VEV of ${\Sigma}$ as
$\langle{\Sigma}\rangle= v_\Phi/\sqrt{2} I$. Under
$SU(2)_{\Phi L}^{\rm global}\times SU(2)_{\Phi R}^{\rm global}$,
the VEV transforms as 
\begin{align}
\langle{\Sigma}\rangle\to
 U_{\Phi L}\langle{\Sigma}\rangle U_{\Phi R}^\dag
 =
 e^{i\theta_{\Phi La}\sigma_a}
 \frac{v_\Phi}{\sqrt{2}} I
 e^{-i\theta_{\Phi Ra}\sigma_a}
 =
 \frac{v_\Phi}{\sqrt{2}} I
 e^{i(\theta_{\Phi La}-\theta_{\Phi Ra})\sigma_a}
 =
 \langle{\Sigma}\rangle
 e^{i(\theta_{\Phi La}-\theta_{\Phi Ra})\sigma_a}.
\end{align}
Therefore, only for $\theta_{\Phi La}=\theta_{\Phi Ra}$, the VEV is
invariant. 
That is, only $SU(2)_{\Phi V}^{\rm global}$ remains.
In the case, a total of three NG or pNG modes appear.

Next, we consider the VEV of $\Delta$ in Eqs.~(\ref{Eq:VEV-full}) and
(\ref{Eq:VEV-reduce}).
When $\kappa=0$, under $SU(2)_{\Delta}^{\rm global}$, the VEV in
Eq.~(\ref{Eq:VEV-reduce}) 
transforms as  
\begin{align}
\langle\Delta\rangle&\to
 U_{\Delta}\langle\Delta\rangle U_{\Delta}^\dag.
\end{align}
For $\theta_{\Delta1}=\theta_{\Delta2}=0$, the VEV is invariant
because $[U_\Delta,\langle\Delta\rangle]=0$.
That is, only $U(1)_{\Delta}^{\rm global}$ associated with
$\sigma_3$ remains.
Under $SU(2)_{\Delta}^{\rm global}$, the VEV in Eq.~(\ref{Eq:VEV-full})
transforms as 
\begin{align}
\langle\Delta\rangle&\to
 U_{\Delta}\langle\Delta\rangle U_{\Delta}^\dag.
\end{align}
When $\kappa\not=0$, 
for $\theta_{\Delta3}=\frac{v_{\eta_3}}{v_{\eta_1}}\theta_{\Delta1}$,
$\theta_{\Delta2}=0$, the VEV is invariant
because $[U_\Delta,\langle\Delta\rangle]=0$.
That is, $U(1)_{\Delta}^{\rm global}$ associated with
a linear combination of $\sigma_1$ and $\sigma_3$ remains. 
From the above discussion, regardless of the configuration of the VEV
of $\Delta$, $SU(2)_{\Delta}^{\rm global}$ symmetry is broken to
$U(1)_{\Delta}^{\rm global}$. Therefore, a total of two NG or pNG modes
appear. 

We summarize NG or pNG modes in this model. When ${\Sigma}(x)$ in
${\bf 2}$ of $SU(2)$ acquires a nonvanishing VEV, a total of three NG or
pNG mode appear. When $\Delta(x)$ in
${\bf 3}$ of $SU(2)$ acquires a nonvanishing VEV, a total of two NG or
pNG mode appear.
In the dark $SU(2)_D$ sector, a total of up to five NG or pNG modes
appear.

\section{Analyzing the potential}
\label{Sec:Potential}

We summarize how to find the vacuum that satisfies the global minimum
of the potential for each set of model parameters below.
\begin{itemize}
 \item[(1)] Write down the most general potential of fields such as
	    $\Sigma(x)$ and $\Delta(x)$. The potential
	    ${\cal V}$ depends on some degrees of freedoms
	    $v_{X}^{}$ such as $v_\Phi$ and $v_{\eta_3}$:
\begin{align}
{\cal V}(\{v_X^{}\})={\cal V}(v_\Phi,v_{\eta_1},v_{\eta_3}). 
\end{align}
	    
 \item[(2)] Calculate the first derivatives of the potential
	    ${\cal V}(\{v_X^{}\})$ with respect to all the 
	    variables $\{v_X^{}\}$.

	    We find three stationary conditions as
\begin{align}
 \frac{\partial}{\partial v_X^{}}{\cal V}(\{v_X^{}\})=0.
\end{align}
	    
 \item[(3)] Solve the simultaneous equations derived from the stationary
	    conditions.

       We find that the variables $v_X^{}$ are expressed as model
       parameters such as $\mu_\Phi^2$ and $\lambda_\Phi$.
       Note that in some cases a VEV is related with another VEV, and
       some flat directions exist. This situation occurs when symmetry
       is unbroken.
       
 \item[(4)] Compare the values of the potential at all extrema and
	    saddle  points.

	    We find true vacua of the potential ${\cal V}(\{v_X^{}\})$
	    at each  parameter region, where all the VEVs at the true
	    vacuum must  be real in our convention.

 \item[(5)] Check what kind of symmetry is realized at each vacuum
	    for each parameter region.
      
\end{itemize}
(Note that the same procedure is commonly used, e.g., to analyze the
vacuum structures of $SU(N)$ symmetry breaking by elementary scalar
fields \cite{Li:1973mq,Meljanac:1982rc} and $E_6$, $SU(N)$ and $SO(N)$
symmetry breaking by composite scalar fields
\cite{Nambu:1961tp,Nambu:1961fr,Kugo:1994qr,Kugo:2019isl,Kugo:2019wge}.)

To understand the vacuum structure of this system, we first consider the
case $\kappa=0$. After that, we will discuss the case $\kappa\not=0$.

\subsection{Without soft symmetry breaking term $(\kappa=0)$}
\label{Seq:Without-soft-breaking}

We take the VEVs of $\Sigma(x)$ and $\Delta(x)$ given in
Eq.~(\ref{Eq:VEV-reduce}). Substituting the VEVs into
the potential of $\Sigma(x)$ and $\Delta(x)$ given in
Eq.~(\ref{Eq:Potential-scalar-check}) with $\kappa=0$
\begin{align}
 {\cal V}({\Sigma},\Delta)
 &=
 -\frac{\mu_\Phi^2}{2}
 \mbox{Tr}\left({\Sigma}^\dag{\Sigma}\right)
 -\frac{1}{2}\mu_{\Delta}^2\mbox{Tr}\left(\Delta^2\right)
 \nonumber\\
 &\hspace{1em}
 +\frac{\lambda_\Phi}{4}
 \left(\mbox{Tr}\left({\Sigma}^\dag{\Sigma}\right)\right)^2
 +\frac{1}{4}\lambda_{\Delta}\mbox{Tr}\left(\Delta^2\right)^2
 +\frac{1}{2}\lambda_{\Phi\Delta}
 \mbox{Tr}\left({\Sigma}^\dag{\Sigma}\right)
  \mbox{Tr}\left(\Delta^2\right),
\label{Eq:Potential-Sigma-Delta-without-kappa}
\end{align}
the potential is given by
\begin{align}
 {\cal V}(v_\Phi,v_\Delta)&=
 -\frac{1}{2}\mu_{\Phi}^2
 v_\Phi^2
 -\frac{1}{2}\mu_{\Delta}^2
 v_\Delta^2
 +\frac{1}{4}\lambda_{\Phi}
 v_\Phi^4
 +\frac{1}{4}\lambda_{\Delta}
 v_\Delta^4
 +\frac{1}{2}\lambda_{\Phi\Delta}
 v_\Phi^2
 v_\Delta^2,
\label{Eq:Potential-Sigma-Delta-without-kappa-VEVs}
\end{align}
where we denote $v_{\eta_3}$ as $v_\Delta$.
This potential is invariant under
$SU(2)_{\Delta}^{\rm global}\times
SU(2)_{\Phi L}^{\rm global}\times SU(2)_{\Phi R}^{\rm global}$
and $SU(2)_D^{\rm local}$
transformations shown in
Figures~\ref{Figure:Global-Symmetry-Breaking-Soft-Breaking-Terms} and
\ref{Figure:Gauge-Symmetry-Breaking}.

Next, we calculate the first derivatives of the potential
${\cal V}(v_\Phi,v_\Delta)$ with respect to
$v_\Phi$ and $v_\Delta$. 
\begin{align}
 \frac{\partial}{\partial v_{\Phi}}
 {\cal V}(v_\Phi,v_\Delta)
 &=v_\Phi
 \left(
 -\mu_{\Phi}^2
 +\lambda_{\Phi}
 v_\Phi^2
 +\lambda_{\Phi\Delta}
 v_\Delta^2
 \right),
 \allowdisplaybreaks[1]\nonumber\\
 \frac{\partial}{\partial v_\Delta}
 {\cal V}(v_\Phi,v_\Delta)
 &=
 v_\Delta
 \left(
 -\mu_{\Delta}^2
 + \lambda_{\Delta}
 v_\Delta^2
 +\lambda_{\Phi\Delta}
 v_\Phi^2
 \right).
\end{align}
From the first derivatives,
we find the following stationary conditions:
\begin{align}
 0&=
v_\Phi
 \left(
 -\mu_{\Phi}^2
 +\lambda_{\Phi}
 v_\Phi^2
 +\lambda_{\Phi\Delta}
 v_\Delta^2
 \right),
 \label{Eq:stationary-condition-1-without-kappa}
 \allowdisplaybreaks[1]\\
 0&=
 v_\Delta
 \left(
 -\mu_{\Delta}^2
 + \lambda_{\Delta}
 v_\Delta
 +\lambda_{\Phi\Delta}
 v_\Phi^2
 \right).
 \label{Eq:stationary-condition-2-without-kappa}
\end{align}

We analytically solve the simultaneous equations given in 
Eqs.~(\ref{Eq:stationary-condition-1-without-kappa}) and
(\ref{Eq:stationary-condition-2-without-kappa}) below.
\begin{itemize}
 \item From Eq.~(\ref{Eq:stationary-condition-1-without-kappa}), we find
       
\begin{align}
 v_\Phi=0\ \ 
 \mbox{or}\ \ 
 -\mu_{\Phi}^2
 +\lambda_{\Phi}
 v_\Phi^2
 +\lambda_{\Phi\Delta}
 v_\Delta^2
 =0.
\label{Eq:VEV-condition-without-kappa}
\end{align}       

 \item First, for $v_\Phi=0$ case,
       from Eqs.~(\ref{Eq:stationary-condition-1-without-kappa}) and
       (\ref{Eq:stationary-condition-2-without-kappa}), we find
\begin{align}
 v_\Delta=0\ \
 \mbox{or}\ \
 v_\Delta=\pm\sqrt{\frac{\mu_\Delta^2}{\lambda_\Delta}}.
\end{align}
       For the first case, $v_\Delta=0$,
       the VEVs are located at the origin
\begin{align}
v_\Phi=v_\Delta=0.
\end{align}
       $SU(2)_D^{\rm local}$ is unbroken, and
       $SU(2)_{\Delta}^{\rm global}\times 
       SU(2)_{\Phi L}^{\rm global}\times SU(2)_{\Phi R}^{\rm global}$ 
       is also unbroken.
       
       For the second case,
       $v_\Delta=\pm\sqrt{\mu_\Delta^2/\lambda_\Delta}$,
       the VEVs are given by
\begin{align}
 v_\Phi=0,\ \ 
 v_\Delta=\pm\sqrt{\frac{\mu_\Delta^2}{\lambda_\Delta}}.
\end{align}
       $SU(2)_D^{\rm local}$ is broken to its subgroup
       $U(1)_D^{\rm local}$, and
       $U(1)_{\Delta}^{\rm global}\times 
       SU(2)_{\Phi L}^{\rm global}\times SU(2)_{\Phi R}^{\rm global}$ 
       remains.
       
 \item Next we consider the second condition in
       Eq.~(\ref{Eq:VEV-condition-without-kappa}). 
       
       From Eq.~(\ref{Eq:stationary-condition-2-without-kappa}), we find
       
\begin{align}
 v_\Delta=0\ \
 \mbox{or}\ \ 
 -\mu_{\Delta}^2
 + \lambda_{\Delta}
 v_\Delta
 +\lambda_{\Phi\Delta}
 v_\Phi^2=0.
\end{align}       

       For the first case $v_\Delta=0$, we find
\begin{align}
 v_\Phi=\pm\sqrt{\frac{\mu_\Phi^2}{\lambda_\Phi}},\ \
 v_\Delta=0. 
\end{align}       
       $SU(2)_D^{\rm local}$ is completely broken.
       $SU(2)_{\Phi L}^{\rm global}\times SU(2)_{\Phi R}^{\rm global}$ 
       is broken to the $SU(2)_{\Phi V}^{\rm global}$ custodial
       symmetry that is the diagonal subgroup of 
       $SU(2)_{\Phi L}^{\rm global}\times SU(2)_{\Phi R}^{\rm global}$, 
       so $SU(2)_{\rm \Delta}^{\rm global}\times
       SU(2)_{\Phi V}^{\rm global}$ remains.
       
 \item Finally, we consider the following simultaneous equations:
\begin{align}
 0=
 -\mu_{\Phi}^2
 +\lambda_{\Phi}
 v_\Phi^2
 +\lambda_{\Phi\Delta}
 v_\Delta^2,\ \ 
 0=
 -\mu_{\Delta}^2
 +\lambda_{\Delta}
 v_\Delta^2
 +\lambda_{\Phi\Delta}
 v_\Phi^2.
\end{align}

       Since the simultaneous equations can be decomposed into two
       quadratic equations of $v_\Phi$ and $v_\Delta$,
       it can be solved as
\begin{align}
 v_\Phi=\pm
 \sqrt{
 \frac{\lambda_\Delta\mu_\Phi^2-\lambda_{\Phi\Delta}\mu_\Delta^2}
 {\lambda_\Delta\lambda_\Phi-\lambda_{\Phi\Delta}^2}},\ \
 v_\Delta=\pm
 \sqrt{
 \frac{\lambda_\Phi\mu_\Delta^2-\lambda_{\Phi\Delta}\mu_\Phi^2}
 {\lambda_\Delta\lambda_\Phi-\lambda_{\Phi\Delta}^2}},
\end{align}
       where all sign combinations exist.
       $SU(2)_D^{\rm local}$ is completely broken.
       $SU(2)_{\Delta}^{\rm global}$
       is broken to $U(1)_{\Delta}^{\rm global}$ and
       $SU(2)_{\Phi L}^{\rm global}\times SU(2)_{\Phi R}^{\rm global}$ 
       is broken to the $SU(2)_{\Phi V}^{\rm global}$ custodial symmetry.
       Therefore
       $U(1)_\Delta^{\rm global}\times SU(2)_{\Phi V}^{\rm global}$
       remains.
       
\end{itemize}

We summarize the extrema and saddle points in the potential
given in Eq.~(\ref{Eq:Potential-Sigma-Delta-without-kappa-VEVs})
in Table~\ref{Tab:Extrema}.  In the table, the potential energy at each
extremum or saddle point,  remaining gauge and global symmetry, and 
a total number of NG modes are also listed, where 
$V_1$, $V_2$, $V_3$, and $V_4$
represent the names of the stationary points and the potential energies
at $SU(2)_D^{\rm local}$, $U(1)_D^{\rm local}$,
$SU(2)_{\Delta}^{\rm global}\times SU(2)_{\Phi V}^{\rm global}$,
and $U(1)_{\Delta}^{\rm global}\times SU(2)_{\Phi V}^{\rm global}$
stationary points, respectively.

\begin{table}[thb]
 \begin{center}
{\footnotesize
\begin{tabular}{|c||c|c|c|c|}\hline
 \rowcolor[gray]{0.8}
 Name
 &$V_1$
 &$V_2$
 &$V_3$
 &$V_4$
 \\ \hline
 $(v_\Phi,v_\Delta)$
 &$\left(0,0\right)$
 &$\left(0,\pm\sqrt{\frac{\mu_\Delta^2}{\lambda_\Delta}}\right)$
 &$\left(\pm\sqrt{\frac{\mu_\Phi^2}{\lambda_\Phi}},0\right)$
 &$\left(\pm\sqrt{
 \frac{\lambda_\Delta\mu_\Phi^2-\lambda_{\Phi\Delta}\mu_\Delta^2}
 {\lambda_\Delta\lambda_\Phi-\lambda_{\Phi\Delta}^2}},
 \pm (\Phi\leftrightarrow\Delta)
 \right)
$\\\hline
 ${\cal V}(v_\Phi,v_\Delta)$
 &$0$
 &$-\frac{\mu_\Delta^4}{4\lambda_\Delta}$
 &$-\frac{\mu_\Phi^4}{4\lambda_\Phi}$
 &$-\frac{\lambda_\Phi\mu_\Delta^4
   -2\lambda_{\Phi\Delta}\mu_\Phi^2\mu_\Delta^2
  +\lambda_\Delta\mu_\Phi^4}
 {4(\lambda_\Delta\lambda_\Phi-\lambda_{\Phi\Delta}^2)}$\\\hline
 $\begin{array}{c}
  \mbox{Gauge}\\
  \mbox{symmetry}\\
  \end{array}$
 &$SU(2)_D$
 &$U(1)_D$
 &None
 &None
 \\\hline
 $\begin{array}{c}
  \mbox{Global}\\
  \mbox{symmetry}\\
   \end{array}$
 &$\begin{array}{c}
  SU(2)_\Delta\\
  \times SU(2)_{\Phi L}\\
  \times SU(2)_{\Phi R}\\
 \end{array}$
 &$\begin{array}{c}
  U(1)_\Delta\\
  \times SU(2)_{\Phi L}\\
  \times SU(2)_{\Phi R}\\
 \end{array}$
 &$\begin{array}{c}
  SU(2)_\Delta\\
  \times SU(2)_{\Phi V}\\
 \end{array}$
 &$\begin{array}{c}
  U(1)_\Delta
  \times SU(2)_{\Phi V}\\
 \end{array}$
 \\\hline
 $\#$ of NG
 &$0$
 &$2$
 &$3$
 &$5$
 \\\hline
\end{tabular}}
 \caption{\small
 The extrema and saddle points in the potential
  given in Eq.~(\ref{Eq:Potential-Sigma-Delta-without-kappa-VEVs})
  for $\kappa=0$ are shown.
  The potential energy at each extremum or saddle point and
  remaining gauge and global symmetry are also listed.
  $\#$ of NG represents the total number of NG modes.
  In the table, the superscript, local/global, is omitted. 
 }
\label{Tab:Extrema}
 \end{center}  
\end{table}

Next, we consider the correspondence between the parameter domain
and the symmetry realized in the vacuum.
First of all, the quartic coupling constants $\lambda_\Delta$,
$\lambda_\Phi$, and $\lambda_{\Phi\Delta}$ must satisfy the following
conditions  to stabilize the potential ${\cal V}(v_\Phi,v_\Delta)$ with
finite values of the VEVs: 
\begin{align}
 \lambda_\Delta>0,\ \ \lambda_\Phi>0,\ \ \
 \lambda_\Delta\lambda_\Phi-\lambda_{\Phi\Delta}^2>0.
\label{Eq:Condition-stability} 
\end{align}
There are four stationary points $V_{1,2,3,4}$ given in
Table~\ref{Tab:Extrema}. They are not always solutions in all parameter
regions because the VEVs $v_\Phi$ and $v_\Delta$ are defined as real
numbers. In fact, $V_1$ is a solution in any $\mu_\Phi^2$ and
$\mu_\Delta^2$ region; $V_2$ is a solution for $\mu_\Delta^2>0$;
$V_3$ is a solution for $\mu_\Phi^2>0$; and 
$V_4$ is a solution for $\mu_\Delta^2>0$ and $\mu_\Phi^2>0$.

\begin{figure}[tbh]
\begin{center}
\includegraphics[bb=0 0 844 216,height=4cm]{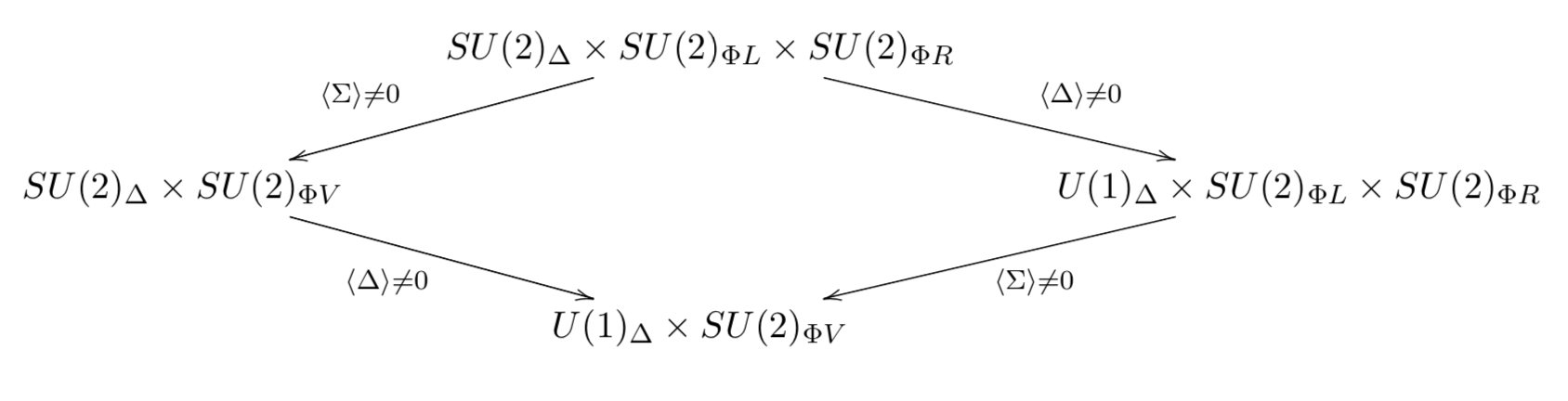}
\end{center}
 \caption{\small The global symmetry
 $SU(2)_\Delta^{\rm global}\times
 SU(2)_{\Phi L}^{\rm global}\times SU(2)_{\Phi R}^{\rm global}$
 breaking patterns for $\kappa=0$
 are shown. 
 $\langle\Sigma\rangle\not=0$ and $\langle\Delta\rangle\not=0$ represent
 the SSB by the VEV of $\Sigma$ and $\Delta$ in
 $({\bf 1,2,2})$ and $({\bf 3,1,1})$ of
 $SU(2)_\Delta^{\rm global}\times
 SU(2)_{\Phi L}^{\rm global}\times SU(2)_{\Phi R}^{\rm global}$,
 respectively.
 In the figure, the superscript, global, is omitted. 
}
\label{Figure:Global-Symmetry-Breaking+Gauge+without-kappa}
\end{figure}

We will find the true vacuum by comparing the potential energies of
stationary points.
When $\mu_\Delta^2>0$ and $\mu_\Phi^2>0$, 
the potential energy preserving
$U(1)_{\Delta}^{\rm global}\times SU(2)_{\Phi V}$
is lower than the other potential energies
preserving $SU(2)_D^{\rm local}$, $U(1)_D^{\rm local}$, and 
$SU(2)_{\Delta}^{\rm global}\times SU(2)_{\Phi V}^{\rm global}$ because
\begin{align}
 V_4-V_1&=-
 \left\{
 \frac{\lambda_\Phi}{4(\lambda_\Delta\lambda_\Phi-\lambda_{\Phi\Delta}^2)}
 \left(\mu_\Delta^2
 -\frac{\lambda_{\Phi\Delta}}{\lambda_\Phi}\mu_\Phi^2\right)^2
 +\frac{\mu_\Phi^4}{4\lambda_\Phi}
 \right\}
 <0,\\
 V_4-V_2&=-
 \frac{(\lambda_\Delta\mu_\Phi^2-\lambda_{\Phi\Delta}\mu_\Delta^2)^2}
 {4\lambda_\Delta(\lambda_\Delta\lambda_\Phi-\lambda_{\Phi\Delta}^2)}
 <0,\\
 V_4-V_3&=-
 \frac{(\lambda_\Phi\mu_\Delta^2-\lambda_{\Phi\Delta}\mu_\Phi^2)^2}
 {4\lambda_\Phi(\lambda_\Delta\lambda_\Phi-\lambda_{\Phi\Delta}^2)}
 <0
\end{align}
from Eq.~(\ref{Eq:Condition-stability}).
For the other parameter spaces of $(\mu_\Phi^2,\mu_\Delta^2)$,
it is easy to find that 
for $\mu_\Phi^2<0$ and $\mu_\Delta^2<0$, $SU(2)_D^{\rm local}$
is realized at the vacuum;
for $\mu_\Phi^2<0$ and $\mu_\Delta^2>0$, $U(1)_{D}^{\rm local}$
is realized at the vacuum;
for $\mu_\Phi^2>0$ and $\mu_\Delta^2<0$,
$SU(2)_{\Delta}^{\rm global}\times SU(2)_{\Phi V}^{\rm global}$ 
is realized at the vacuum.
The global symmetry
$SU(2)_\Delta^{\rm global}\times
SU(2)_{\Phi L}^{\rm global}\times SU(2)_{\Phi R}^{\rm global}$
breaking patterns are shown in
Figure~\ref{Figure:Global-Symmetry-Breaking+Gauge+without-kappa}.

\subsection{With soft symmetry breaking term $(\kappa\not=0)$}

We will now begin analyzing the potential for the case $\kappa\not=0$.
We take the VEVs of $\Sigma(x)$ and $\Delta(x)$ given in
Eq.~(\ref{Eq:VEV-full}). Substituting the VEVs into
the potential of $\Sigma(x)$ and $\Delta(x)$ given in
Eq.~(\ref{Eq:Potential-scalar-check})
\begin{align}
 {\cal V}({\Sigma},\Delta)
 &=
 -\frac{\mu_\Phi^2}{2}
 \mbox{Tr}\left({\Sigma}^\dag{\Sigma}\right)
 -\frac{1}{2}\mu_{\Delta}^2\mbox{Tr}\left(\Delta^2\right)
 -\sqrt{2}\kappa
 \mbox{Tr}
 \left(
 \sigma_3
 {\Sigma}^\dag \Delta{\Sigma}
 \right)
 \nonumber\\
 &\hspace{1em}
 +\frac{\lambda_\Phi}{4}
 \left(\mbox{Tr}\left({\Sigma}^\dag{\Sigma}\right)\right)^2
 +\frac{1}{4}\lambda_{\Delta}\mbox{Tr}\left(\Delta^2\right)^2
 +\frac{1}{2}\lambda_{\Phi\Delta}
 \mbox{Tr}\left({\Sigma}^\dag{\Sigma}\right)
  \mbox{Tr}\left(\Delta^2\right),
\label{Eq:Potential-Sigma-Delta}
\end{align}
the potential is given by
\begin{align}
 {\cal V}(v_\Phi,v_{\eta_1},v_{\eta_3})&=
 -\frac{1}{2}\mu_{\Phi}^2
 v_\Phi^2
 -\frac{1}{2}\mu_{\Delta}^2
 \left(v_{\eta_1}^2+v_{\eta_3}^2\right)
 -\kappa
 v_{\Phi}^2v_{\eta_3}
 \nonumber\\
 &\hspace{1em}
 +\frac{1}{4}\lambda_{\Phi}
 v_\Phi^4
 +\frac{1}{4}\lambda_{\Delta}
 \left(v_{\eta_1}^2+v_{\eta_3}^2\right)^2
 +\frac{1}{2}\lambda_{\Phi\Delta}
 v_\Phi^2
  \left(v_{\eta_1}^2+v_{\eta_3}^2\right).
\label{Eq:Potential-Sigma-Delta-with-kappa-VEVs}
\end{align}
We recall that the $\kappa$ term is invariant under
$SU(2)_{D}^{\rm local}$, but breaks 
$SU(2)_\Delta^{\rm global}\times
SU(2)_{\Phi L}^{\rm global}\times SU(2)_{\Phi R}^{\rm global}$
to $SU(2)_{L}^{\rm global}\times U(1)_{R}^{\rm global}$.
Therefore, this system is invariant under $SU(2)_{D}^{\rm local}$ gauge 
and $SU(2)_{L}^{\rm global}\times U(1)_{R}^{\rm global}$ global
transformations.
Note that $\Sigma(x)$ and $\Delta(x)$ belong to ${\bf 2}(\pm1)$ and
${\bf 3}(0)$ of $SU(2)_{L}^{\rm global}\times U(1)_{R}^{\rm global}$,
respectively, where the numbers in boldface denote $SU(2)_{L}^{\rm
global}$ representation and numbers in parentheses denote
$U(1)_{R}^{\rm global}$ charges.

Next, we calculate the first derivatives of the potential
${\cal V}(v_\Phi,v_{\eta_1},v_{\eta_3})$ with respect to
$v_\Phi,v_{\eta_1},v_{\eta_3}$. 
\begin{align}
 \frac{\partial}{\partial v_{\Phi}}
 {\cal V}(v_\Phi,v_{\eta_1},v_{\eta_3})
 &=v_\Phi
 \left(
 -\mu_{\Phi}^2
 -2\kappa
 v_{\eta_3}
 +\lambda_{\Phi}
 v_\Phi^2
 +\lambda_{\Phi\Delta}
 \left(v_{\eta_1}^2+v_{\eta_3}^2\right)
 \right),
 \allowdisplaybreaks[1]\nonumber\\
 \frac{\partial}{\partial v_{\eta_1}}
 {\cal V}(v_\Phi,v_{\eta_1},v_{\eta_3})
 &=
 v_{\eta_1}
 \left(
 -\mu_{\Delta}^2
 + \lambda_{\Delta}
 \left(v_{\eta_1}^2+v_{\eta_3}^2\right)
 +\lambda_{\Phi\Delta}
 v_\Phi^2
 \right),
 \allowdisplaybreaks[1]\nonumber\\
 \frac{\partial}{\partial v_{\eta_3}}
 {\cal V}(v_\Phi,v_{\eta_1},v_{\eta_3})
 &=
 v_{\eta_3}
 \left(
 -\mu_{\Delta}^2
 +\lambda_{\Delta}
 \left(v_{\eta_1}^2+v_{\eta_3}^2\right)
 +\lambda_{\Phi\Delta}
 v_\Phi^2
 \right)
  -\kappa
 v_{\Phi}^2.
\end{align}
From the first derivatives,
we find the following stationary conditions:
\begin{align}
 0&=
v_\Phi
 \left(
 -\mu_{\Phi}^2
 -2\kappa
 v_{\eta_3}
 +\lambda_{\Phi}
 v_\Phi^2
 +\lambda_{\Phi\Delta}
 \left(v_{\eta_1}^2+v_{\eta_3}^2\right)
 \right),
 \label{Eq:stationary-condition-1}
 \allowdisplaybreaks[1]\\
 0&=
 v_{\eta_1}
 \left(
 -\mu_{\Delta}^2
 + \lambda_{\Delta}
 \left(v_{\eta_1}^2+v_{\eta_3}^2\right)
 +\lambda_{\Phi\Delta}
 v_\Phi^2
 \right),
 \label{Eq:stationary-condition-2}
 \allowdisplaybreaks[1]\\
 0&=
 v_{\eta_3}
 \left(
 -\mu_{\Delta}^2
 +\lambda_{\Delta}
 \left(v_{\eta_1}^2+v_{\eta_3}^2\right)
 +\lambda_{\Phi\Delta}
 v_\Phi^2
 \right)
  -\kappa
 v_{\Phi}^2.
 \label{Eq:stationary-condition-3}
\end{align} 

We analytically solve the simultaneous equations given by
Eqs.~(\ref{Eq:stationary-condition-1}),
(\ref{Eq:stationary-condition-2}), and 
(\ref{Eq:stationary-condition-3}) below.
\begin{itemize}
 \item From Eq.~(\ref{Eq:stationary-condition-1}), we find
\begin{align}
v_\Phi=0\ \ \mbox{or}\ \ 
 -\mu_{\Phi}^2
 -2\kappa
 v_{\eta_3}
 +\lambda_{\Phi}
 v_\Phi^2
 +\lambda_{\Phi\Delta}
 \left(v_{\eta_1}^2+v_{\eta_3}^2\right)=0.
\label{Eq:VEV-condition}
\end{align}       

 \item First, for $v_\Phi=0$ case,
       from Eqs.~(\ref{Eq:stationary-condition-1}),
       (\ref{Eq:stationary-condition-2}), and
       (\ref{Eq:stationary-condition-3}), we find
\begin{align}
 v_{\eta_1}=v_{\eta_3}=0\ \ \mbox{or}\ \
 -\mu_{\Delta}^2
 +\lambda_{\Delta}
 \left(v_{\eta_1}^2+v_{\eta_3}^2\right)^2
 =0.
\end{align}
       For the first case, the VEVs are located at the origin
\begin{align}
v_\Phi=v_{\eta_1}=v_{\eta_3}=0.
\end{align}
       $SU(2)_D^{\rm local}$ is unbroken, and
       $SU(2)_{L}^{\rm global}\times U(1)_{R}^{\rm global}$ 
       is also unbroken.
       
       For the second case, we can take the following VEVs by using 
       the $SU(2)_{L}^{\rm global}$ transformation:
\begin{align}
 v_\Phi=0,\ \ v_{\eta_1}=0,\ \
 v_{\eta_3}=\pm\sqrt{\frac{\mu_\Delta^2}{\lambda_\Delta}}.
\end{align}
       $SU(2)_D^{\rm local}$ is broken to its subgroup
       $U(1)_D^{\rm local}$, and
       $U(1)_{L}^{\rm global}\times U(1)_{R}^{\rm global}$ 
       remains.
       
 \item Next we consider the second condition in
       Eq.~(\ref{Eq:VEV-condition}).       
       From Eq.~(\ref{Eq:stationary-condition-2}), we find
\begin{align}
v_{\eta_1}=0\ \ \mbox{or}\ \
 -\mu_{\Delta}^2
 + \lambda_{\Delta}
 \left(v_{\eta_1}^2+v_{\eta_3}^2\right)
 +\lambda_{\Phi\Delta}
 v_\Phi^2=0.
\end{align}       
       From Eq.~(\ref{Eq:stationary-condition-3}), 
       the above second condition leads to $\kappa=0$, but due to
       $\kappa\not=0$, $v_{\eta_1}=0$.
       For $v_\Phi\not=0$ and $v_{\eta_1}=0$, we need to
       solve the following simultaneous equations:
\begin{align}
 0&=
 -\mu_{\Phi}^2
 -2\kappa
 v_{\eta_3}
 +\lambda_{\Phi}
 v_\Phi^2
 +\lambda_{\Phi\Delta}
 v_{\eta_3}^2,
 \label{Eq:stationary-condition-1'}
 \\
 0&=
 v_{\eta_3}
 \left(
 -\mu_{\Delta}^2
 +\lambda_{\Delta}
 v_{\eta_3}^2
 +\lambda_{\Phi\Delta}
 v_\Phi^2
 \right)
  -\kappa
 v_{\Phi}^2.
 \label{Eq:stationary-condition-3'}
\end{align}
       The solutions of the simultaneous equations lead to
       $v_\Phi\not=0$ and $v_{\eta_3}\not=0$, so
       the vacuum of these solutions breaks 
       $SU(2)_{L}^{\rm global}\times U(1)_{R}^{\rm global}$ to
       $U(1)_{V}^{\rm global}$.

       The simultaneous equations in
       Eqs.~(\ref{Eq:stationary-condition-1'}) and
       (\ref{Eq:stationary-condition-3'}) can be decomposed into a cubic
       equations for $v_{\eta_3}$  and a quadratic equation for
       $v_\Phi$.
       From the vacuum solutions listed in Table~\ref{Tab:Extrema} and
       the soft symmetry breaking $\kappa$ term,
       the three solutions of the cubic equation correspond to one
       $SU(2)_\Delta^{\rm global}\times SU(2)_{\Phi V}^{\rm global}$
       and two
       $U(1)_\Delta^{\rm global}\times SU(2)_{\Phi V}^{\rm global}$
       global symmetry vacuum solutions in the $\kappa=0$ case.
       We can solve the exact solutions of the simultaneous equations
       because of just cubic and quadratic equations,
       but they are too complicated to show here.
       Instead, we can find approximate solutions to the simultaneous
       equations by using the solutions around the
       $SU(2)_\Delta^{\rm global}\times SU(2)_{\Phi V}^{\rm global}$
       and 
       $U(1)_\Delta^{\rm global}\times SU(2)_{\Phi V}^{\rm global}$       
       vacuum solutions.  The detailed values are not important for the
       discussion here, so we omit the particular form, but there are
       solutions around the $\kappa=0$ solution
       in Table~\ref{Tab:Extrema} as follows.
       From the solution for the
       $SU(2)_\Delta^{\rm global}\times SU(2)_{\Phi V}^{\rm global}$
       vacuum  in the $\kappa=0$ case,
\begin{align} 
 v_\Phi=\pm\sqrt{\frac{\mu_\Phi^2}{\lambda_\Phi}}+O(\kappa),\ \ 
 v_{\eta_3}=O(\kappa).
\end{align}       
       From the solution for the
       $U(1)_\Delta^{\rm global}\times SU(2)_{\Phi V}^{\rm global}$       
       vacuum in the $\kappa=0$ case,
\begin{align}
 v_\Phi=\pm\sqrt{
 \frac{\lambda_\Delta\mu_\Phi^2-\lambda_{\Phi\Delta}\mu_\Delta^2}
 {\lambda_\Delta\lambda_\Phi-\lambda_{\Phi\Delta}^2}}
 +O(\kappa),\ \ 
 v_{\eta_3}= 
 \pm
\sqrt{
 \frac{\lambda_\Phi\mu_\Delta^2-\lambda_{\Phi\Delta}\mu_\Phi^2}
 {\lambda_\Delta\lambda_\Phi-\lambda_{\Phi\Delta}^2}}
 +O(\kappa).
\label{Eq:VEVs-with-kappa-U1V}
\end{align}       
       
\end{itemize}

\begin{table}[thb]
 \begin{center}
{\footnotesize
\begin{tabular}{|c||c|c|c|c|}\hline
 \rowcolor[gray]{0.8}
 Name
 &$V_{\rm I}$
 &$V_{\rm II}$
 &$V_{\rm III}$
 &$V_{\rm IV}$
 \\ \hline
 $(v_\Phi,v_{\eta_3})$
 &$\left(0,0\right)$
 &$\left(0,\pm\sqrt{\frac{\mu_\Delta^2}{\lambda_\Delta}}\right)$
 &$\left(\pm\sqrt{\frac{\mu_\Phi^2}{\lambda_\Phi}},
 O(\kappa)\right)$
 &$\left(\pm\sqrt{
 \frac{\lambda_\Delta\mu_\Phi^2-\lambda_{\Phi\Delta}\mu_\Delta^2}
 {\lambda_\Delta\lambda_\Phi-\lambda_{\Phi\Delta}^2}},
 \pm (\Phi\leftrightarrow\Delta)
 \right)
$\\\hline
 ${\cal V}(v_\Phi,v_{\eta_3})$
 &$0$
 &$-\frac{\mu_\Delta^4}{4\lambda_\Delta}$
 &$-\frac{\mu_\Phi^4}{4\lambda_\Phi}$
 &$-\frac{\lambda_\Phi\mu_\Delta^4
   -2\lambda_{\Phi\Delta}\mu_\Phi^2\mu_\Delta^2
  +\lambda_\Delta\mu_\Phi^4}
 {4(\lambda_\Delta\lambda_\Phi-\lambda_{\Phi\Delta}^2)}$\\\hline
 $\begin{array}{c}
  \mbox{Gauge}\\
  \mbox{symmetry}\\
  \end{array}$
 &$SU(2)_D$
 &$U(1)_D$
 &\multicolumn{2}{c|}{None}\\\hline
 $\begin{array}{c}
  \mbox{Global}\\
  \mbox{symmetry}\\
   \end{array}$
 &$SU(2)_{L}\times U(1)_{R}$
 &$U(1)_{L}\times U(1)_{R}$
 &\multicolumn{2}{c|}{$U(1)_{V}$}\\\hline
 $\#$ of NG
 &$0$
 &$2$
 &$3$
 &$5$
 \\\hline
\end{tabular}}
 \caption{\small
 The extrema and saddle points in the potential
  given in Eq.~(\ref{Eq:Potential-Sigma-Delta-with-kappa-VEVs})
  for $\kappa\not=0$ are shown.
  The potential energy at each extremum or saddle point and
  remaining gauge and global symmetry are also listed, where 
  $v_{\eta_1}=0$.
  $\#$ of NG represents the total number of NG and pNGB modes.
  In the table, the superscript, local/global, is omitted. 
  For $U(1)_{V}^{\rm global}$ case, we omit $O(\kappa)$ if there is
  already a value greater than $\kappa$. 
 }
\label{Tab:Extrema+}
 \end{center}  
\end{table}

\begin{figure}[tbh]
\begin{center}
\includegraphics[bb=0 0 638 215,height=4cm]{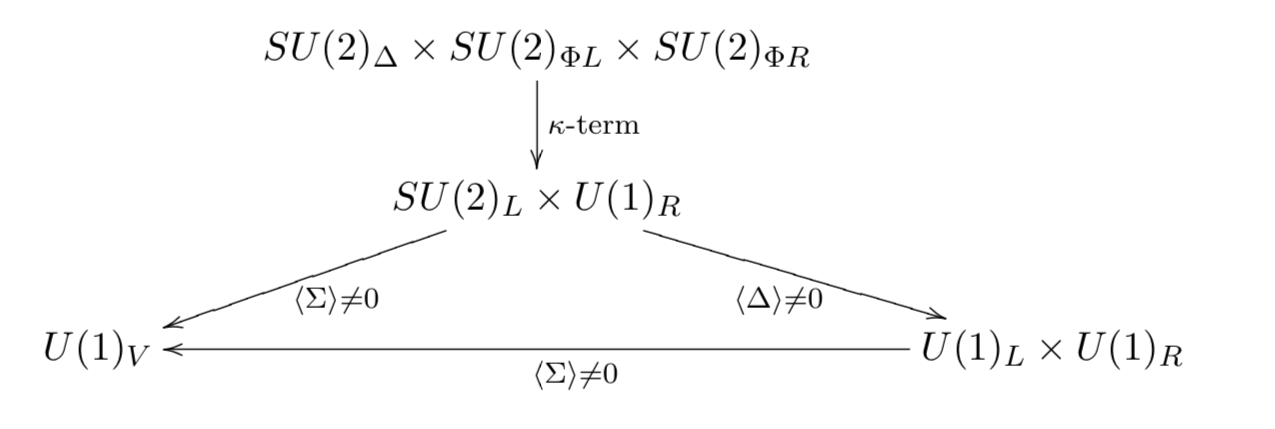}
 \end{center}
 \caption{\small The global symmetry
 $SU(2)_\Delta^{\rm global}\times
 SU(2)_{\Phi L}^{\rm global}\times SU(2)_{\Phi R}^{\rm global}$
 breaking patterns for $\kappa\not=0$ are shown.
 The $\kappa$ term stands for a soft symmetry  breaking term;
 $\langle\Sigma\rangle\not=0$ and $\langle\Delta\rangle\not=0$ represent
 the SSB by the VEVs of $\Sigma$ and $\Delta$ in
 $({\bf 1,2,2})$ and $({\bf 3,1,1})$ of
 $SU(2)_\Delta^{\rm global}\times
 SU(2)_{\Phi L}^{\rm global}\times SU(2)_{\Phi R}^{\rm global}$.
 $({\bf 1,2,2})$ and $({\bf 3,1,1})$ of
 $SU(2)_\Delta^{\rm global}\times
 SU(2)_{\Phi L}^{\rm global}\times SU(2)_{\Phi R}^{\rm global}$
 are decomposed into two ${\bf 2}$ and one ${\bf 3}$ of
 $SU(2)_L$, respectively.
 In the figure, the superscript, global, is omitted. 
}
\label{Figure:Global-Symmetry-Breaking+Gauge}
\end{figure}

Next, we consider the correspondence between the parameter domain
and the symmetry realized in the vacuum. Since the $\kappa$ term does
not affect the shape of the potential at infinity, the constraint of the
parameter region from the stability condition to potential is the same
for $\kappa\not=0$  as for $\kappa=0$, which is given in
Eq.~(\ref{Eq:Condition-stability}).
In the region where $\kappa$ can be treated perturbatively, the true
vacuum does not change, so the results for the case $\kappa=0$ are
applicable.
Therefore, 
for ${}^\forall\mu_\Phi^2$ and $\mu_\Delta^2>0$,
$U(1)_{V}^{\rm global}$
is realized at the vacuum;
for $\mu_\Phi^2<0$ and $\mu_\Delta^2<0$, $SU(2)_D^{\rm local}$
is realized at the vacuum;
for $\mu_\Phi^2<0$ and $\mu_\Delta^2>0$, $U(1)_D^{\rm local}$
is realized at the vacuum.
The extrema and saddle points in the potential are summarized in 
Table~\ref{Tab:Extrema+}. 
The global symmetry
$SU(2)_\Delta^{\rm global}\times
SU(2)_{\Phi L}^{\rm global}\times SU(2)_{\Phi R}^{\rm global}$
breaking patterns are shown in
Figure~\ref{Figure:Global-Symmetry-Breaking+Gauge}.

Before investigating the mass spectra of the scalar sectors, we comment
on the would-be NG and pNG modes.
For $V_{\rm I}$ of Table~\ref{Tab:Extrema+}, there are no NG modes.
For $V_{\rm II}$, there are two NG modes, and they are absorbed by
the $SU(2)_D^{\rm local}/U(1)_D^{\rm local}$ gauge bosons.
For $V_{\rm III}$, there are three NG modes, and they
are absorbed by the $SU(2)_D^{\rm local}$ gauge bosons.
For $V_{\rm IV}$, there are five NG modes. Three of the five NG modes
are absorbed in the $SU(2)_D^{\rm local}$ gauge boson. The remaining two
NG modes are real scalar modes with $U(1)_V^{\rm global}$ charges and
are identified as one complex scalar.

As can be seen from the above discussion, a charged pNGB that can be
regarded as a DM appears only when $\mu_\Phi^2>0$ and $\mu_\Delta^2>0$.
In the following, we consider such a case.

\section{Mass spectrum}
\label{Sec:Mass-spectrum}

Here we investigate the mass spectra of the scalar fields $\Sigma(x)$
(or $\Phi(x)$), $\Delta(x)$, and $H(x)$ for the parameter region
$\mu_\Phi^2>0$ and $\mu_\Delta^2>0$, where $U(1)_{V}^{\rm global}$
symmetry is realized at a vacuum. In particular, we confirm that
there is a $U(1)_{V}^{\rm global}$ charged complex scalar with a
mass proportional to the $\kappa$ parameter. It corresponds to the pNG
mode, which will be regarded as a DM candidate.

First, we check the potential terms associated with $\Sigma$ and
$\Delta$. Here we consider the following field expression at the
vacuum: 
\begin{align}
 \Delta(x)
 =\frac{1}{\sqrt{2}}\left(
 \begin{array}{cc}
   v_\Delta+\eta_3
   &\eta_1-i\eta_2 \\
   \eta_1+i\eta_2
   &-v_\Delta-\eta_3\\
 \end{array}
 \right),\ \ \
 \Sigma(x)
=\frac{1}{\sqrt{2}}\left(
 \begin{array}{cc}
  v_\Phi+\phi_3-i\phi_4&\phi_1+i\phi_2\\
   \phi_1-i\phi_2& v_\Phi+\phi_3+i\phi_4\\
 \end{array}
 \right),
 \label{Eq:VEV-Delta-Phi}
\end{align}
where we denote the VEV of $\eta_3$ as $v_\Delta$, and 
the values of $v_\Delta$ and $v_\Phi$ for $\kappa\to 0$ are
given in Eq.~(\ref{Eq:VEVs-with-kappa-U1V}). In our convention,
the generator of the remaining $U(1)_V^{\rm global}$ corresponds to
the $\sigma_3$ direction. Therefore, $\eta_1$, $\eta_2$, $\phi_1$, and
$\phi_2$ have the same $U(1)_V^{\rm global}$ charge, while
$\eta_3$, $\phi_3$, and $\phi_4$ have no $U(1)_V^{\rm global}$ charge.
The $U(1)_V^{\rm global}$ charge of fields can be checked by
using, e.g., the generator of the $U(1)_V^{\rm global}$ charge
$\sigma_3$. 

The stationary conditions in
Eqs.~(\ref{Eq:stationary-condition-1'}) and
(\ref{Eq:stationary-condition-3'}) for $v_\Phi,v_\Delta\not=0$
can be written as
\begin{align}
 \mu_{\Phi}^2&=-2\kappa
 v_\Delta
 +\lambda_{\Phi}
 v_\Phi^2
 +\lambda_{\Phi\Delta}
 v_{\eta_3}^2,
 \label{Eq:Mu-Phi}
 \\
\mu_{\Delta}^2&=
 \lambda_{\Delta}
 v_\Delta^2
 +\lambda_{\Phi\Delta}
 v_\Phi^2
 -\kappa
 \frac{v_{\Phi}^2}{v_\Delta},
\label{Eq:Mu-Delta}
\end{align}
where we replaced $v_{\eta_3}$ as $v_\Delta$.

Substituting Eqs.~(\ref{Eq:VEV-Delta-Phi}), (\ref{Eq:Mu-Phi}), and
(\ref{Eq:Mu-Delta}) into the potential ${\cal V}(\Sigma,\Delta)$ given in 
Eq.~(\ref{Eq:Potential-Sigma-Delta}),
the potential is given as
\begin{align}
 {\cal V}(\eta_1,\eta_2,\eta_3,\phi_1,\phi_2,\phi_3,\phi_4)
={\cal V}_0+{\cal V}_2+{\cal V}_3+{\cal V}_4,
\end{align}
where ${\cal V}$ stands for $V_{\rm IV}$ in Table~\ref{Tab:Extrema+},
and the subscript of ${\cal V}_j (j=0,2,3,4)$ denotes the mass dimension
of the operator.
Note that the tadpole term of the potential ${\cal V}_1$ disappears from
the stationary conditions. The constant terms of the potential of the real
scalar fields
$(\eta_1,\eta_2,\eta_3,\phi_1,\phi_2,\phi_3,\phi_4)$ are given by
\begin{align}
 {\cal V}_0=-\frac{1}{4}\left(
 \lambda_\Delta v_\Delta^4
 +\lambda_\Phi v_\Phi^4
 +2\lambda_{\Phi\Delta}v_\Delta^2v_\Phi^2
 -2\kappa v_\Delta v_\Phi^2\right).
\label{Eq:Dim-0-term}
\end{align}
The quadratic terms are given by
\begin{align}
 {\cal V}_{2}&=
 \frac{1}{2}
  \left(
 \begin{array}{cc}
  \phi_1&\eta_1\\
 \end{array}
 \right) 
 \left(
 \begin{array}{cc}
  4\kappa v_\Delta&2\kappa v_\Phi\\
  2\kappa v_\Phi&\frac{v_\Phi^2}{v_\Delta}\kappa\\
 \end{array}
 \right) 
 \left(
 \begin{array}{c}
  \phi_1\\
  \eta_1\\
 \end{array}
 \right)
 +\frac{1}{2}
 \left(
 \begin{array}{cc}
  \phi_2&\eta_2\\
 \end{array}
 \right) 
 \left(
 \begin{array}{cc}
  4\kappa v_\Delta&-2\kappa v_\Phi\\
  -2\kappa v_\Phi&\frac{v_\Phi^2}{v_\Delta}\kappa\\
 \end{array}
 \right) 
 \left(
 \begin{array}{c}
  \phi_2\\
  \eta_2\\
 \end{array}
 \right)
 \nonumber\\
 &\hspace{1em}
 +\frac{1}{2}
 \left(
 \begin{array}{cc}
  \phi_3&\eta_3\\
 \end{array}
 \right) 
 \left(
 \begin{array}{cc}
 2\lambda_\Phi v_\Phi^2
   &2\lambda_{\Phi\Delta}v_\Phi v_\Delta
   -2v_\Phi\kappa\\
  2\lambda_{\Phi\Delta}v_\Phi v_\Delta
   -2v_\Phi\kappa
   & 2\lambda_\Delta v_\Delta^2+\frac{v_\Phi^2}{v_\Delta}\kappa\\
 \end{array}
 \right) 
 \left(
 \begin{array}{c}
  \phi_3\\
  \eta_3\\
 \end{array}
 \right)
 +0\times \phi_4^2.
\label{Eq:Dim-2-term}
\end{align}
Since the determinant of $(\eta_1,\phi_1)$ is zero, one of the
eigenvalues is zero. Furthermore, if $\kappa$ is set to zero, both
eigenvalues are zero. The same is true for $(\eta_2,\phi_2)$.
Since the determinant of $(\eta_3,\phi_3)$ is non-zero, even for
$\kappa\to 0$ zero eigenvalues do not appear. Instead of it, 
a $U(1)_V^{\rm global}$ neutral scalar field $\phi_4$ is always
massless. Therefore, we find that one of the two linear combinations of
$\eta_1$ and $\phi_1$ is an NG mode that is absorbed by an
$SU(2)_D^{\rm local}$ gauge boson, and the other is a pNG mode.
The same is true for $\eta_2$ and $\phi_2$. $\phi_4$ is an NG mode, and
$\eta_3$ and $\phi_3$ are Higgs modes.
Similarly, the cubic and quartic terms are given by
\begin{align}
 {\cal V}_3
  &=2\kappa\left(\eta_1\phi_1\phi_3
 -\eta_2\phi_2\phi_3+\eta_2\phi_1\phi_4
 +\eta_1\phi_2\phi_4\right)
 +\kappa \eta_3
 \left(\phi_1^2+\phi_2^2-\phi_3^2-\phi_4^2\right)
 \nonumber\\
 &+ \lambda_\Delta v_\Delta \eta_3
 \left(\eta_1^2+\eta_2^2+\eta_3^2\right)
 +\lambda_\Phi v_\Phi\phi_3
 \left(\phi_1^2+\phi_2^2+\phi_3^2+\phi_4^2\right)
 \nonumber\\
 &+\lambda_{\Phi\Delta}v_\Delta \eta_3
 \left(\phi_1^2+\phi_2^2+\phi_3^2+\phi_4^2\right)
 +\lambda_{\Phi\Delta}v_\Phi\phi_3
 \left(\eta_1^2+\eta_2^2+\eta_3^2\right),
\label{Eq:Dim-3-term}
 \\
 {\cal V}_4
 &=\frac{1}{4}\lambda_\Delta
 \left(\eta_1^2+\eta_2^2+\eta_3^2\right)^2
 +\frac{1}{4}\lambda_\Phi
 \left(\phi_1^2+\phi_2^2+\phi_3^2+\phi_4^2\right)^2
 \nonumber\\
 &\ \ +\frac{1}{2}\lambda_{\Phi\Delta}
 \left(\eta_1^2+\eta_2^2+\eta_3^2\right)
 \left(\phi_1^2+\phi_2^2+\phi_3^2+\phi_4^2\right).
\label{Eq:Dim-4-term}
\end{align}

For the calculation of the scattering amplitudes of SM particles and a
charged pNGB in the next section, here we switch on the scalar field $H$
in ${\bf 2}$ of $SU(2)_W^{\rm local}$ listed in
Table~\ref{Tab:Matter_content} and rewrite charged scalar fields as
follows:
\begin{align}
 &H=\frac{1}{\sqrt{2}}
 \left(
 \begin{array}{c}
  0\\
  v+h\\
 \end{array}
\right),\ \ \
 \phi_{(\pm)}:=
 \frac{1}{\sqrt{2}}\left(\phi_1\pm i\phi_2\right),\ \ \
 \eta_{(\pm)}:=
 \frac{1}{\sqrt{2}}\left(\eta_1\mp i\eta_2\right),
\label{Eq:Higgs-charged}
\end{align}
where the subscripts (positive and negative signs) enclosed in
parentheses indicate the sign of the $U(1)_V$ charge, and 
$v$ stands for the VEV of the SM Higgs boson, which breaks
$SU(2)_W^{\rm local}\times U(1)_Y^{\rm local}$ into
$U(1)_{\rm EM}^{\rm local}$.
In the rest of this section, we will examine the mass matrix when a
$U(1)_V^{\rm global}$ neutral scalar field $h$ is added to the mass
matrix given in Eq.~(\ref{Eq:Dim-2-term}), where the field $h$ is a main
component of the SM Higgs boson.
 
First, we consider the mass eigenstates for the $U(1)_V^{\rm global}$
charged scalar fields $\phi_{1,2}$ and $\eta_{1,2}$.
When we rewrite $\phi_{1,2}$ and $\eta_{1,2}$ in terms of
$\phi_{(\pm)}$ and $\eta_{(\pm)}$ as in Eq.~(\ref{Eq:Higgs-charged}), the mass
term of $\phi_{1,2}$ and $\eta_{1,2}$ given in Eq.~(\ref{Eq:Dim-2-term})
can be rewritten as 
\begin{align}
{\cal V}_2\ni \frac{1}{2}
  \left(
 \begin{array}{cc}
  \phi_{(\mp)}&\eta_{(\mp)}\\
 \end{array}
 \right) 
 M_C^2
 \left(
 \begin{array}{c}
  \phi_{(\pm)}\\
  \eta_{(\pm)}\\
 \end{array}
 \right),\ \ 
 M_C^2:=
 \left(
 \begin{array}{cc}
  4\kappa v_\Delta&2\kappa v_\Phi\\
  2\kappa v_\Phi&\frac{v_\Phi^2}{v_\Delta}\kappa\\
 \end{array}
 \right).
\label{Eq:Mass-matrix-charged}
\end{align}
The above mass matrix can be easily diagonalized, and the mass and
mixing matrix are given by
\begin{align}
 &m_{G_{(\pm)}}=0,\ \
 m_{\varphi}^2=\kappa\left(4v_\Delta+\frac{v_\Phi^2}{v_\Delta}\right),
 \ \ 
 \left(
 \begin{array}{c}
  G_{(\pm)}\\
  \varphi_{(\pm)}\\
 \end{array}
 \right)
 :=\left(
 \begin{array}{cc}
  \cos\beta&\sin\beta\\
  -\sin\beta&\cos\beta\\
 \end{array}
 \right)
 \left(
 \begin{array}{c}
  \phi_{(\pm)}\\
  \eta_{(\pm)}\\  
 \end{array}
 \right),
\label{Eq:Def-G-DM} 
\end{align}
where 
\begin{align}
 \sin\beta=\frac{v_\Phi}{\sqrt{v_\Phi^2+4v_\Delta^2}},\ \ 
 \cos\beta=\frac{2v_\Delta}{\sqrt{v_\Phi^2+4v_\Delta^2}},
\label{Eq:Def-alpha} 
\end{align}
$G_{(\pm)}$ is the would-be NG modes of $SU(2)_D^{\rm local}$, and 
$\varphi_{(\pm)}$ is a $U(1)_V^{\rm global}$ charged scalar mode that is
a pNG mode of $SU(2)_V^{\rm global}/U(1)_V^{\rm global}$, which will be
identified as a DM.

Second, even when $h$ exists, there is no mass mixing of $\phi_4$ and
$h$, so $\phi_4$ remains massless. That is, $\phi_4$ is a would-be NG
mode of $SU(2)_D^{\rm local}$. Further, $\phi_4$ has no charge of 
$U(1)_D^{\rm local}$ and $U(1)_V^{\rm global}$. Therefore, $\phi_4$ can
be identified with the neutral NG mode $G_0$.

Finally, the mass matrix of a $U(1)_V^{\rm global}$ neutral sector
$(\eta_3,\phi_3,h)$ is given by
\begin{align}
 M_H^2:=\left(
  \begin{array}{ccc}
   2\lambda_H v^2
    &\lambda_{H\Phi}vv_\Phi
    &2\lambda_{H\Delta}vv_\Delta
    \\
    \lambda_{H\Phi}vv_\Phi
    &2\lambda_\Phi v_\Phi^2
    &2\lambda_{\Phi\Delta}v_\Phi v_\Delta
   -2v_\Phi\kappa
   \\
  2\lambda_{H\Delta}vv_\Delta
  &2\lambda_{\Phi\Delta}v_\Phi v_\Delta   -2v_\Phi\kappa
  &    2\lambda_\Delta v_\Delta^2+\frac{v_\Phi^2}{v_\Delta}\kappa
   -2v_\Phi\kappa
  \\
  \end{array}
 \right).
\label{Eq:Mass-matrix-neutral}
\end{align}
Since this mass matrix is a real symmetric matrix, it can be
diagonalized by a unitary matrix (orthogonal matrix) $U_H$, where
$U_H^\dag U_H=I$ and $U_H^\dag=U_H^T$. That is,
$U_H M_H^2 U_H^\dag=(M_H^2)^{\rm diag}$, where $(M_H^2)^{\rm diag}$ is a
$3\times 3$
diagonal matrix. The mass eigenstates can be expressed from
the original basis as follows: 
\begin{align}
 \left(
 \begin{array}{c}
  h_1\\
  h_2\\
  h_3\\
 \end{array}
 \right)
 :=U_H
 \left(
 \begin{array}{c}
  h\\
  \phi_3\\
  \eta_3\\
 \end{array}
 \right),
\end{align}
where $h_{j}(j=1,2,3)$ are mass eigenstates with no $U(1)_V^{\rm global}$
charge, and $h_1$ is identified as the observed SM Higgs mode with a
mass of about 125\,GeV.
The exact eigenvalues and eigenvectors are too complicated to show here.
Instead of it, we show the approximate mass eigenvalues and mass mixing
matrix when $v_\Delta$ is sufficiently larger than $v$ and $v_\Phi$. For
$v_\Delta\gg v_\Phi,v$, the mixing matrix $U_H$ is given by 
\begin{align}
 U_H&=
 \left(
 \begin{array}{ccc}
  1&0&-\frac{\lambda_{H\Delta}v}{\lambda_\Delta v_\Delta} \\
   0&1&-\frac{\lambda_{\Phi\Delta}v_\Phi}{\lambda_\Delta v_\Delta}\\
  \frac{\lambda_{H\Delta}v}{\lambda_\Delta v_\Delta}
   &\frac{\lambda_{\Phi\Delta}v_\Phi}{\lambda_\Delta v_\Delta}&1\\
 \end{array}
 \right)
 \left(
 \begin{array}{ccc}
  \cos\alpha&\sin\alpha&0 \\
  -\sin\alpha&\cos\alpha&0\\
  0&0&1\\
 \end{array}
 \right)+O\left(\frac{v^2}{v_\Delta^2},\frac{v_\Phi^2}{v_\Delta^2}\right),
 \nonumber\\
 &\tan2\alpha\simeq
 \frac{2vv_\Phi\left(
 \lambda_{H\Phi}\lambda_\Phi
 -\lambda_{H\Delta}\lambda_{\Phi\Delta}\right)}
 {v^2\left(\lambda_{H\Delta}^2-\lambda_H\lambda_\Delta\right)
 -v_\Phi^2(\lambda_{S\Phi}^2-\lambda_\Phi\lambda_\Delta)}.
\label{Eq:UH-approximate}
\end{align}
The mass eigenvalues for $h_i$ are given by
\begin{align}
 m_{h_1}^2&\simeq
 \lambda_Hv^2
 -\frac{\lambda_{H\Delta}^2\lambda_\Phi-2\lambda_{H\Phi}
 \lambda_{H\Delta}\lambda_{\Phi\Delta}
 +\lambda_\Delta\lambda_{H\Phi}^2}
 {\lambda_{\Phi}\lambda_\Delta-\lambda_{\Phi\Delta}^2}v_\Phi^2,
 \nonumber\\
 m_{h_2}^2&\simeq
 \frac{\lambda_\Phi\lambda_\Delta-\lambda_{\Phi\Delta}^2}
 {\lambda_\Delta}v_\Phi^2
 +\frac{\left(\lambda_\Delta\lambda_{H\Phi}
 -\lambda_{H\Delta}\lambda_{\Phi\Delta}\right)^2}
 {\lambda_\Delta(\lambda_\Phi\lambda_\Delta)
 -\lambda_{\Delta\Phi}^2}v^2,
 \nonumber\\
 m_{h_3}^2&\simeq
 \lambda_\Delta v_\Delta^2.
\end{align}       

Here we comment on the mass matrix in
Eq.~(\ref{Eq:Mass-matrix-neutral}). The similar mass matrix has been
analyzed in $G_{\rm SM}\times U(1)_{B-L}$ and $SO(10)$ pNGB DM
models Refs.~\cite{Abe:2020iph,Okada:2020zxo,Abe:2021byq,Okada:2021qmi},
but only the case $v_\Delta\gg v_\Phi,v$ such as
$v_\Delta> O(10^{10})$\,GeV and $v_\Phi,v=O(10^2)$\,GeV is allowed in
those models due to the DM stability problem. In this model, the
stability of the DM is guaranteed by $U(1)$, but when we identify
$\varphi_{(\pm)}$ as the DM, the direct detection leads to some constraints,
which we will discuss in the next section.

\section{Direct detection and relic abundance}
\label{Sec:Scattering}

\begin{figure}[tbh]
\begin{center}
\includegraphics[bb=0 0 442 292,height=6cm]{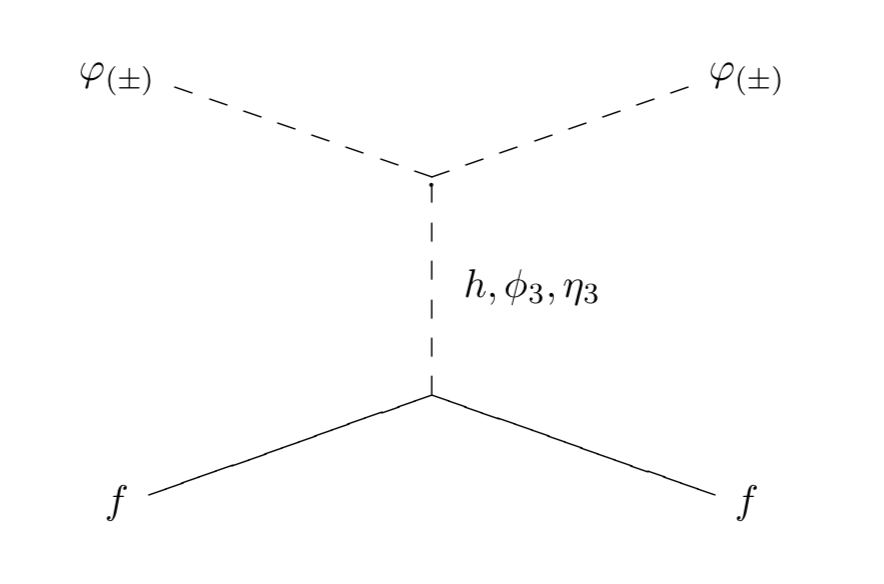}
\end{center}
 \caption{\small Tree-level scattering of $\varphi_{(\pm)}$ and the SM
 fermions $f$ via the scalar fields $h,\phi_3,\eta_3$ is shown.
}
\label{Figure:Scattering}
\end{figure}

In this section, we will show how the model introduced in Sec.~\ref{Sec:Model}
is constrained by various DM experiments.
Firstly, we study the scattering amplitudes of a DM candidate
$\varphi_{(\pm)}$ and SM fermions via the SM Higgs and additional
scalar fields shown in Figure~\ref{Figure:Scattering}. 
In the original pNGB DM model (Ref.~\cite{Gross:2017dan} for Abelian and
Ref.~\cite{Abe:2022mlc} for non-Abelian case), the soft breaking term is a
scalar bilinear term and gives only an origin of pNGB DM mass term.  
Such soft breaking terms preserve a nature of NGB for DM and gives
derivative portal interactions, resulting in vanishing DM-nucleon 
scattering amplitudes in $t\to 0$ limit.
In the model we introduced in Sec.~\ref{Sec:Model}, however, the soft breaking term found in
Eq.~(\ref{Eq:Potential-Sigma-Delta}) is a scalar trilinear term, and
gives not only an origin of pNGB DM mass term but also additional 
$h_i \varphi_{(+)}\varphi_{(-)}$ portal interactions proportional to soft
breaking parameter $\kappa$, just like Refs.~\cite{Abe:2020iph} and
\cite{Abe:2021byq}. These portal interactions give rise to new
contribution in addition to canceling diagrams, resulting in a nonzero
DM-nucleon scattering process even in $t\to 0$ limit. Therefore, we must
look into parameter regions that escape direct detection constraints. 

Apart from the vanishing part in $t\to 0$ limit, the DM-nucleon
scattering amplitude  $\mathcal{A}_{dd}$ in our model is proportional to
soft breaking parameter $\kappa$, which is replaced with 
the DM mass $m_\varphi$ by using Eq.~(\ref{Eq:Def-G-DM}) as
\begin{align}
{\cal A}_{dd}&\underset{\sim}{\propto}
\frac{m_\varphi^2}{4v_\Delta^2}
\frac{m_f}{v}
\left[
v_\Phi\sin2\alpha_z
\left(
\frac{1}{m_{h_1}^2}-\frac{1}{m_{h_2}^2}
\right)
+
4v_\Delta
\left(
-\frac{1}{m_{h_1}^2}\alpha_y \cos \alpha_z
+\frac{1}{m_{h_2}^2}\alpha_x \sin \alpha_z
\right)
\right] ,
\label{Eq:Add-Cross-Section} 
\end{align}
where $m_f$ denotes the mass of SM fermions $f$;
$\alpha_x$, $\alpha_y$, and $\alpha_z$ stand for the mixing angles of
$\phi_3\mathchar`-\eta_3$, $h\mathchar`-\eta_3$, and $h\mathchar`-\phi_3$, respectively;
\begin{align}
\left(  \begin{array}{c}
         h_1
         \\
         h_2
         \\
         h_3
         \end{array}
\right)
=
\left(  \begin{array}{ccc}
         1 & 0 & 0
         \\
         0 & \cos\alpha_x & \sin\alpha_x 
         \\
         0 & -\sin\alpha_x & \cos\alpha_x 
         \end{array}
\right)  
\left(  \begin{array}{ccc}
         \cos\alpha_y & 0 &  \sin\alpha_y
         \\
         0 & 1 & 0 
         \\
         -\sin\alpha_y & 0 &  \cos\alpha_y
         \end{array}
\right)  
\left(  \begin{array}{ccc}
         \cos\alpha_z & \sin\alpha_z & 0
         \\
         -\sin\alpha_z & \cos\alpha_z & 0
         \\
         0 & 0 & 1 
         \end{array}
\right)         
\left(  \begin{array}{c}
         h
         \\
         \phi_3
         \\
         \eta_3
         \end{array}
\right).
\end{align}
Note that the mixing angles $\alpha_x$ and $\alpha_y$ are expressed in terms 
of VEVs of the scalar fields and four point interaction coefficients as
\begin{align}
\alpha_x \simeq -\frac{\lambda_{H\Delta} v}{\lambda_\Delta v_\Delta},
\quad
\alpha_y \simeq -\frac{\lambda_{\Phi\Delta}v_\Phi}{\lambda_\Delta v_\Delta}.
\end{align}
Note also that we retain only the first order term for $v/v_\Delta$ or $v_\Phi/v_\Delta$ in Eq.~(\ref{Eq:Add-Cross-Section}).
See Appendix~\ref{Sec:DM-quark-scattering} for the detailed derivation.

As commented in the previous section, the previous models such as 
$G_{\rm SM}\times U(1)_{B-L}$ and $SO(10)$ pNGB DM models 
required a very high $v_\Delta$ due to DM longevity.
The high $v_\Delta$ also brings about a small DM-nucleon scattering amplitude,
because it is suppressed by $1/v_\Delta^2$.
In this model, on the other hand, the stability of the DM is guaranteed
by $U(1)_V$, so $v_\Delta$ is expected to be allowed to be a much smaller
scale than  $O(10^{10})$\,GeV, as we will see later.

In the remainder of this section, we compare the spin-independent (SI)
DM-nucleon cross section $\sigma_{\rm SI}$
and show limitations on $v_\Delta$ from recent DM experiments
\cite{XENON:2018voc,LUX-ZEPLIN:2022qhg}.
In the model, the SI cross section
$\sigma_{\rm SI}$ is approximately given by
\begin{align}
 \sigma_{\rm SI}&\simeq
 \frac{1}{16\pi}
 \Bigg(
\frac{m_\varphi^2}{2v_\Delta^2}
\frac{m_f}{v}
\Bigg[
v_\Phi\sin2\alpha_z
\left(
\frac{1}{m_{h_1}^2}-\frac{1}{m_{h_2}^2}
\right)
\nonumber\\
&\qquad\qquad\qquad\qquad
+
4v_\Delta
\left(
-\frac{1}{m_{h_1}^2}\alpha_y \cos \alpha_z
+\frac{1}{m_{h_2}^2}\alpha_x \sin \alpha_z
\right)
\Bigg]
 \Bigg)^2
 \frac{m_p^{4}f_N^2}{\left(m_\varphi+m_p\right)^2},
\label{Eq:SI-cross-section} 
\end{align}
where the proton mass $m_p\simeq 0.938$\,GeV, and $f_N\simeq 0.3$.
Note that we retain only the first order term for $v/v_\Delta$ or $v_\Phi/v_\Delta$ in Eq.~(\ref{Eq:SI-cross-section}).
For the conversion formula from DM-quark scattering to DM-nucleon scattering, 
see e.g., Ref.~\cite{Arcadi:2019lka}.

The thermal relic abundance of DM in a model that can be regarded as a
low-energy effective description of this model has been calculated in
Ref.~\cite{Abe:2022mlc}, and it has been shown that the observed value
can be reproduced when the DM mass satisfies the condition
$m_{\rm DM}\gtrsim m_{h_1}/2$. Furthermore, there is a constraint from
Higgs invisible decay when the DM mass is less than half of the Higgs
boson mass. Therefore, in the following, we mainly focus on regions
where the DM mass is more than half the Higgs boson mass:
$m_\varphi\gtrsim m_{h_1}/2$.

\begin{figure}[htb]
\begin{center}
\includegraphics[bb=0 0 636 595,height=9cm]{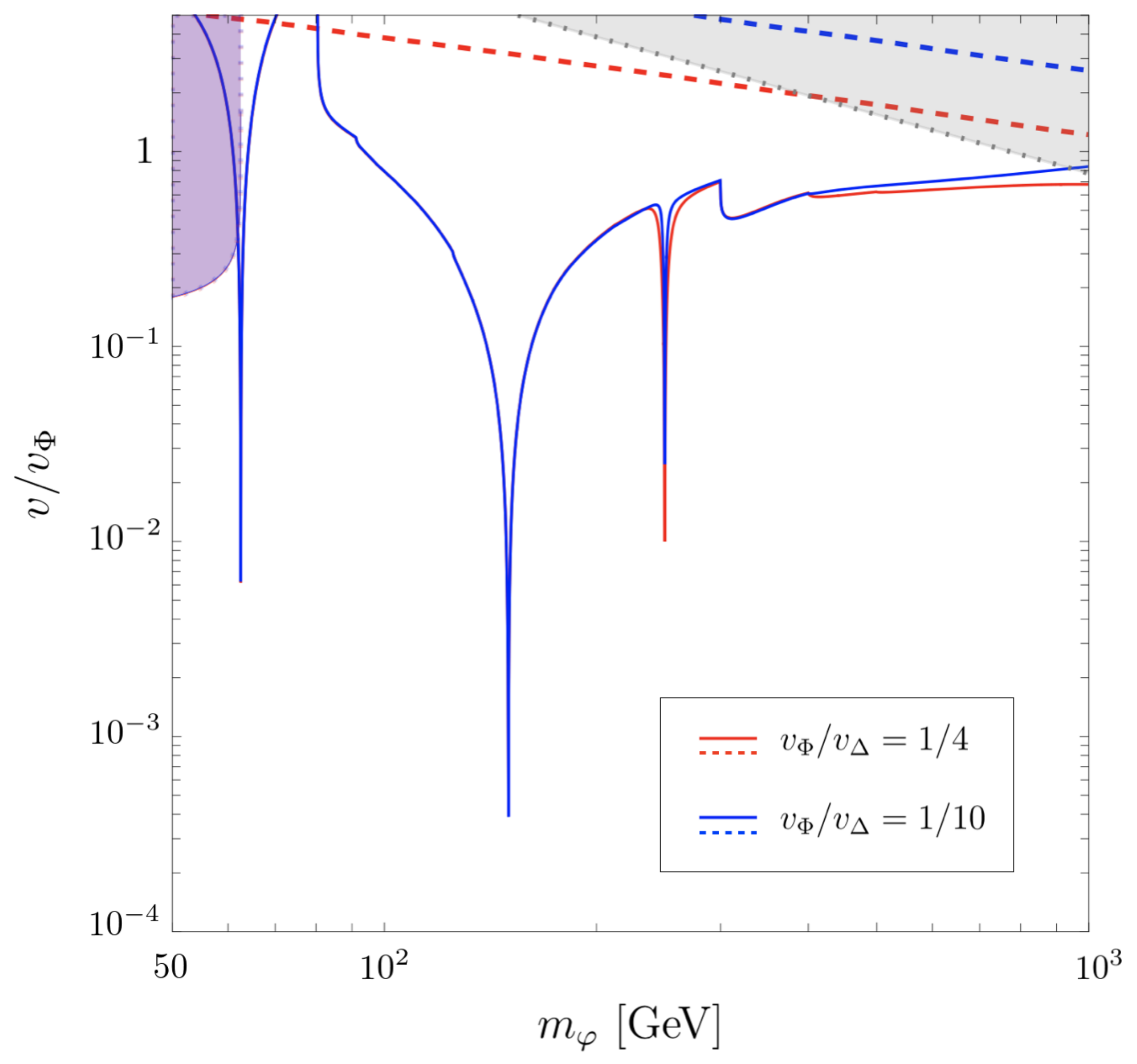}
 \caption{\small 
Constraints from a direct detection experiment and a prediction for the
relic abundance in our $SU(2)$ pNGB DM model. 
Solid lines express a parameter contour corresponding to 
$\Omega h^2 =  0.12$, while dashed line represents a direct detection
 constraints from the LUX-ZEPLIN experiment\cite{LUX-ZEPLIN:2022qhg}.
 The gray shaded region is the region where the $SU(2)_D$ neutral gauge boson can also become DM candidate, which is not dealt with in this analysis for simplicity.
 The purple shaded region is excluded by Higgs invisible decay constraints\cite{ATLAS:2020kdi}.
 }
\label{Fig:dd_vs_relic}
\end{center}
\end{figure}

As a benchmark parameter set, we fix mass parameters for the second and 
third neutral Higgs fields as $m_{h_2}=300$\,GeV and $m_{h_3}=500$\,GeV,
respectively. We take a sample set
$(\sin\alpha_x,\,\sin\alpha_y,\,\sin\alpha_z)=(0.06,0.05,0.1)$. 
We assume that $SU(2)_D$ gauge bosons are heavy.

In Figure~\ref{Fig:dd_vs_relic}, we show allowed parameter regions 
consistent with various experimental constraints, varying the ratio of
the $SU(2)_D$ doublet and triplet VEV as $v_\Phi/v_\Delta = 1/4$ and
$1/10$. Solid lines express parameter contours reproducing an observed
DM energy density, $\Omega h^2 =  0.12$.
Three dips in solid lines correspond to the resonance contributions from 
$h_1$, $h_2$, and $h_3$. Therefore, they locate at the half of their
masses, $m_{h_1}/2=$62.5\,GeV, $m_{h_2}/2=$150\,GeV, and
$m_{h_3}/2=$250\,GeV, 
respectively. The small depression at $m_\varphi=300$\,GeV is due to the
opening of a new annihilation channel, $\varphi_{(+)}\varphi_{(-)} \to h_2h_2$. 
Dashed lines represent constraints from a direct detection LUX-ZEPLIN
experiment\cite{LUX-ZEPLIN:2022qhg} recasted to the upper limit  
for the VEV ratio $v/v_\Phi$. The direct detection constraints become
tight at large DM mass region. That is because the DM-nucleon scattering
amplitude is proportional to a soft-breaking parameter, namely, the DM
mass square $m_\varphi^2$, as we show in
Eq.~(\ref{Eq:Add-Cross-Section}). The gray shaded region satisfies
$m_{Z'}<2m_\varphi$ with $m_{Z'}$ being mass of the $SU(2)_D$ neutral
gauge boson. In this region, the $SU(2)_D$ neutral gauge boson can also
become DM candidate, which is not dealt with in this analysis for
simplicity. The purple shaded region is excluded by Higgs invisible
decay constraints\cite{ATLAS:2020kdi}. The Higgs invisible decay width
in this model shows $v_\Phi/v_\Delta$-dependence only through
sub-leading terms. Therefore, the excluded region colored in purple 
is common for VEV ratio $v_\Phi/v_\Delta = 1/4$ and $1/10$. 

We apply the same method in
Refs.~\cite{Arcadi:2017kky,Arcadi:2019lka,Arcadi:2021mag} to calculate
the DM-nucleon scattering cross-sections and thermally averaged total
annihilation cross sections. 
We find that the relic abundance does not change so much
when we vary the VEV ratio $v_\Phi/v_\Delta$. 
We also find that there are plenty of allowed parameter
regions which escape the direct detection constraints and reproduce a
correct DM relic abundance at the same time.

\section{Summary and discussions}
\label{Sec:Summary}

We proposed a new pNGB DM model based on non-Abelian gauge symmetry
$SU(2)_D$, in which scalars in ${\bf 2}$ and ${\bf 3}$ of $SU(2)_D$ are
introduced.
We analyzed the structure of the symmetry and its breaking
patterns in detail by analyzing the scalar potential. We found that when
the mass parameters of the scalars $\mu_\Phi^2$ and $\mu_\Delta^2$ are
positive in our convention, the $SU(2)_D$ gauge symmetry is
spontaneously broken to the exact $U(1)_V$ global symmetry by the VEVs
of the scalars in ${\bf 2}$ and ${\bf 3}$ of $SU(2)_D$.
The charged pNGB under the $U(1)_V$ custodial symmetry appears, and is
identified as DM. The stability of the DM is guaranteed by the exact
$U(1)_V$ custodial symmetry. From Figure~\ref{Fig:dd_vs_relic},
we showed that the relic abundance is correctly reproduced while escaping
the severe constraints from the direct detection experiments.

We comment on the additional $SU(2)_D$ gauge symmetry breaking scale.
In the $G_{\rm SM}\times U(1)_{B-L}$ and $SO(10)$ pNGB DM models
\cite{Abe:2020iph,Okada:2020zxo,Abe:2021byq,Okada:2021qmi},
the VEV of the additional $U(1)_{B-L}$ gauge symmetry breaking scale
must be higher than ${\cal O}(10^{10})$\,GeV to suppress DM decay rate,
while in our new pNGB DM model the VEV of $SU(2)_D$ breaking scale is 
allowed to be roughly ${\cal O}(1)$\,TeV or higher due to the stability
of DM guaranteed by the $U(1)_V$ custodial symmetry.
Complementary verification by accelerator experiments may be possible in
some parameter regions in our model.

In the $SU(2)_g\times U(1)_X$ pNGB DM model
\cite{Abe:2022mlc}, it has been pointed out that an additional gauge
coupling of $U(1)_X$ is not asymptotically free and there is a Landau
pole in the high-energy region, so this problem can be tackled by
extending to $SU(2)$ gauge theory.
However, it is not enough to extend the additional gauge sector part to
non-Abelian gauge symmetry because the SM gauge group includes an
Abelian gauge symmetry $U(1)_Y$.
To address this issue, we have to discuss extensions to grand unified
theory (GUT)
\cite{Georgi:1974sy,Slansky:1981yr,Yamatsu:2015gut}. The extension of
this model to GUT models will be left as future work.

\section*{Acknowledgments}

This work was supported in part
by the MEXT Grant-in-Aid for Scientific Research on Innovation Areas
Grant No. JP18H05543 (T.S., K.T., and Y.U.),
JSPS Grant-in-Aid for Scientific Research KAKENHI Grant Nos.
JP20K14477 (H.O.),
JP18K03651 (T.S.), JP18H01210 (T.S.), JP22K03622 (T.S.),
JP22K03620 (K.T.),
the Ministry of Science and Technology of Taiwan under
Grant No. MOST-111-2811-M-002-047-MY2 (N.Y.),
and the Education and Research Program for Mathematical and Data Science
from the Kyushu University (H.O.).

\appendix

\section{DM-quark scattering amplitude}
\label{Sec:DM-quark-scattering}

We will derive DM-quark scattering amplitude given in 
Eq.~(\ref{Eq:Add-Cross-Section}).
The scalar kinetic terms $\mathcal{K}$ and scalar potential
$\mathcal{V}$ parts of the Lagrangian in Eq.~(\ref{Eq:Lagrangian})
are given as 
\begin{align}
\label{eq:model_K}
\mathcal{K}
=&~
{(D_\mu H)^\dagger D^\mu H}
+\frac{1}{2} \mbox{Tr}
\left({(D_\mu \Sigma)^\dagger D^\mu \Sigma}\right)
+\frac{1}{2} \mbox{Tr}
\left({D_\mu \Delta D^\mu \Delta}\right),\\
\label{eq:model_V}
\mathcal{V}
=&~
{-\mu_H^2H^\dag H}
-\frac{1}{2}\mu_\Phi^2\mbox{Tr}
\left({\Sigma^\dagger \Sigma}\right)
-\frac{1}{2}\mu_\Delta^2\mbox{Tr}
\left({\Delta^2}\right)
-\sqrt{2}\kappa \mbox{Tr}
\left({\sigma_3 \Sigma^\dagger \Delta \Sigma}\right)
\nonumber\\
&~
{+\lambda_{H}\left(H^\dag H\right)^2}
+\frac{\lambda_\Phi}{4} \left( \mbox{Tr}
\left({\Sigma^\dagger \Sigma}\right) \right)^2
+\frac{\lambda_\Delta}{4} \left( \mbox{Tr}
\left({\Delta}^2\right) \right)^2
\nonumber\\
&~
 {
 +\frac{\lambda_{H\Phi}}{2}
 \left(H^\dag H\right)
 \mbox{Tr}\left({\Sigma}^\dag{\Sigma}\right)
 +\frac{\lambda_{H\Delta}}{2}\left(H^\dag H\right)
 \mbox{Tr}\left(\Delta^2\right)
 }
+\frac{\lambda_{\Phi\Delta}}{2}
 \mbox{Tr}
 \left({\Sigma^\dagger \Sigma}\right)\mbox{Tr}
 \left({\Delta^2}\right),
\end{align}
where $\Sigma=(\tilde{\Phi}, \Phi)$ and $\Delta$ are
$SU(2)_D^{\rm local}$ bi-doublet and real triplet fields, respectively.

For deriving relevant interactions for DM-quark scattering amplitude, we
adopt non-linear basis, because in the non-linear basis, the
Higgs-portal interactions only come from  kinetic terms in
Eq.~(\ref{eq:model_K}) and the soft breaking term in the first line of 
Eq.~(\ref{eq:model_V}), which makes derivation of DM-quark scattering
amplitude much easier than that in the linear basis. 
Note that the result Eq.~(\ref{Eq:Add-Cross-Section}) is the same
regardless of whether we choose a linear or no-linear basis.
The polar decomposition for bi-doublet and real triplet fields 
are given as
\begin{align}
\label{eq:model_sigma-polar}
&\Sigma
=
\frac{v_{\Phi}+\phi_3}{\sqrt{2}}
{\xi} _{\Sigma} 
\qquad\qquad\quad
\mbox{with}
\qquad
{\xi}_{\Sigma}
=
\exp\left( \frac{i}{v_\Phi}(\phi_{(+)}\sigma_{(+)} + \phi_{(-)} \sigma_{(-)} + \phi_4
 \sigma_3) \right),
\\
\label{eq:model_delta-polar}
&\Delta
=
\frac{v_\Delta + \eta_3}{\sqrt{2}} 
{\xi}_{\Delta}^\dagger
\sigma_3
{\xi}_{\Delta} 
\qquad~\,
\mbox{with}
\qquad
{\xi}_{\Delta} 
=
\exp \left(-\frac{i}{2v_\Delta}(\eta_{(+)}\sigma_{(+)} + \eta_{(-)} \sigma_{(-)} )\right),
\end{align}
where $v_\Phi$, and $v_\Delta$ are VEVs for each scalar field. 
$\sigma_{(\pm)}$ is expressed in terms of the first and second Pauli
matrices as $\sigma_{(\pm)} = ({\sigma_1 \mp i \sigma_2})/{\sqrt{2}}$.
Note that $SU(2)$ generators $\sigma_{(\pm)}$, $\sigma_3$ satisfy 
the normalization conditions 
$\mbox{Tr}\left({\sigma_{(+)}\sigma_{(+)}}\right)
=\mbox{Tr}\left({\sigma_{(-)}\sigma_{(-)}}\right)=0$,
$\mbox{Tr}\left({\sigma_{(+)} \sigma_{(-)}}\right)
=\mbox{Tr}\left({\sigma_3\sigma_3}\right) = 2$,
and bi-doublet and real triplet fields satisfy
\begin{align}
\mbox{Tr}\left({\Sigma^\dagger \Sigma}\right) = (v_\Phi + \phi_3)^2 \, ,
\qquad
\mbox{Tr}\left({\Delta^2}\right) = (v_\Delta + \eta_3)^2 \, .
\end{align}

We evaluate scattering amplitudes of the DM $\varphi_{{(\pm)}}$ and the SM
fermions $f$ shown in Figure~\ref{Figure:Scattering}. Substituting polar
decompositions
Eqs.~(\ref{eq:model_sigma-polar}) and (\ref{eq:model_delta-polar})  
into the potential $\mathcal{V}$ and kinetic terms $\mathcal{K}$
given in Eq.~(\ref{eq:model_K}) and Eq.~(\ref{eq:model_V}), 
and extracting cubic scalar interactions relevant to DM-fermion
scattering, we get
\begin{align}
\label{eq:amp_Vnl-3p}
&-\mathcal{V}_{\rm NL}
\supset~
-\frac{2}{v_\Phi} \phi_3~   
(\phi_{(-)}, \eta_{(-)})
M_C^2
    \left(  \begin{array}{c}
         \phi_{(+)}
         \\
         \eta_{(+)}
         \end{array}
    \right)        
-\frac{1}{v_\Delta} \eta_3~   
(\phi_{(-)}, \eta_{(-)})
M_C^2
    \left(  \begin{array}{c}
         \phi_{(+)}
         \\
         \eta_{(+)}
         \end{array}
 \right),
 \\
\label{eq:amp_Knl-3p}
&\mathcal{K}_{\rm NL}
\supset~
\left( 1+ \frac{\phi_3}{v_\Phi} \right)^2 \partial_\mu \phi_{(+)} \partial^\mu \phi_{(-)}
+\left( 1 + \frac{\eta_3}{v_\Delta} \right)^2 \partial_\mu \eta_{(+)} \partial^\mu \eta_{(-)} \, .
\end{align}
where $M_C^2$ is given in Eq.~(\ref{Eq:Mass-matrix-charged}).
The cubic interactions in Eq.~(\ref{eq:amp_Vnl-3p}) can be 
rewritten in terms of mass eigenstate $\varphi_{(\pm)}$ as
\begin{align}
\label{eq:amp_Vnl-mass}
-\mathcal{V}_{\rm NL}
\supset&~
\left( -2\frac{m_\varphi^2}{v_\Phi}\phi_3 
    - \frac{m_\varphi^2}{v_\Delta}\eta_3 
\right)  
\varphi_{(+)}\varphi_{(-)} \, .
\end{align}
The cubic interactions in Eq.~(\ref{eq:amp_Knl-3p}) can be 
rewritten in terms of $\varphi_{(\pm)}$ as
\begin{align}
\label{eq:amp_Knl-mass}
\mathcal{K}_{\rm NL}
&\supset
2
\left(  \frac{\sin^2\beta}{v_\Phi} \phi_3
     + \frac{\cos^2\beta}{v_\Delta} \eta_3
\right)
\partial_\mu \varphi_{(+)} \partial^\mu \varphi_{(-)}
\nonumber\\
&=
\left(  \frac{2}{v_\Phi} \phi_3
     + \frac{1}{v_\Delta} \eta_3
\right)
\partial_\mu \varphi_{(+)} \partial^\mu \varphi_{(-)}
-
\left(  2\,\frac{v_\Phi}{v_\Phi^2 + 4 v_\Delta^2} \phi_3 
     + \frac{1}{v_\Delta} \frac{4v_\Delta^2 - v_\Phi^2}{v_\Phi^2 + 4 v_\Delta^2} \eta_3
\right)
\partial_\mu \varphi_{(+)} \partial^\mu \varphi_{(-)} \, .
\end{align}
Combining the first term in the last line of Eq.~(\ref{eq:amp_Knl-mass}) 
with Eq.~(\ref{eq:amp_Vnl-mass}), we get
\footnote{
{
To obtain Eq.~(\ref{eq:amp_Vnl+Knl}), we used
\begin{align}
\rho\, \partial_\mu \varphi_{(+)} \partial^\mu \varphi_{(-)}
=&~
\frac{1}{2}\partial_\mu
  \Big[  \rho\, (\partial^\mu \varphi_{(+)})\varphi_{(-)}
           + \rho \,\varphi_{(+)}(\partial^\mu \varphi_{(-)})
           - (\partial^\mu \rho) \varphi_{(+)}\varphi_{(-)}
  \Big]
\nonumber\\
&
+\frac{1}{2}(\partial^2 \rho)\varphi_{(+)}\varphi_{(-)}
-\frac{1}{2}\rho\, (\partial^2 \varphi_{(+)}) \varphi_{(-)}
-\frac{1}{2}\rho\, \varphi_{(+)} (\partial^2 \varphi_{(-)})
\label{Eq:Integration-by-parts}
\end{align}
with $\rho = \phi_3, \eta_3$. 
Total derivative terms in the first line of Eq.~(\ref{Eq:Integration-by-parts}) 
are irrelevant, so we dropped them in Eq.~(\ref{eq:amp_Vnl+Knl}). 
}
}
\begin{align}
\label{eq:amp_Vnl+Knl}
&\left(  \frac{2}{v_\Phi} \phi_3
     + \frac{1}{v_\Delta} \eta_3
\right)
\partial_\mu \varphi_{(+)} \partial^\mu \varphi_{(-)}
-
\left( 2\frac{m_\varphi^2}{v_\Phi}\phi_3 
    + \frac{m_\varphi^2}{v_\Delta}\eta_3 
\right) 
\varphi_{(+)}\varphi_{(+)}
\nonumber\\
&
 =
\left(  \frac{2}{v_\Phi} \phi_3
     + \frac{1}{v_\Delta} \eta_3
\right)
\bigg[
-\frac{1}{2}  \Big(  [\partial^2 + m_\varphi^2]\, \varphi_{(+)} \Big) \varphi_{(-)}
-\frac{1}{2} \varphi_{(+)} \Big( [\partial^2 + m_\varphi^2]\, \varphi_{(-)} \Big)
\bigg]
\nonumber\\
&\ \ \
+\left( \frac{2}{v_\Phi}(\partial^2 \phi_3) + \frac{1}{v_\Delta}(\partial^2 \eta_3)\right) \varphi_{(+)}\varphi_{(-)}.
\end{align}
We find that all the terms appearing in Eq.~(\ref{eq:amp_Vnl+Knl}) are
irrelevant to the DM-fermion scattering in direct detection;
the first line vanishes due to on-shell conditions for pNGB DM;
the second line gives contributions proportional to 
momentum-transfer $t=(p_2-p_1)^2$ with $p_{1,2}$ being in-coming and 
out-going DM momentum. This will also vanish when we take $t\to 0$ limit.

The remaining cubic interaction relevant to DM-fermion scattering
shown in Figure~\ref{Figure:Scattering} is the second term of
Eq.~(\ref{eq:amp_Knl-mass}):
\begin{align}
\mathcal{K}_{\rm NL}
\supset&~
-
\left(  2\,\frac{v_\Phi}{v_\Phi^2 + 4 v_\Delta^2} \phi_3 
+ \frac{1}{v_\Delta} \frac{4v_\Delta^2 - v_\Phi^2}{v_\Phi^2 + 4
 v_\Delta^2} \eta_3
\right)
 \partial_\mu \varphi_{(+)} \partial^\mu \varphi_{(-)}.
\end{align}
Note that this term decouples when we assume $v_\Delta \gg v_\Phi$.
Further, by using the relation in Eq.~(\ref{Eq:Integration-by-parts}),
we obtain
\begin{align}
\mathcal{L}_{\rm NL}
\supset&~
-{\frac{1}{2}}
\left(  2\,\frac{v_\Phi}{v_\Phi^2 + 4 v_\Delta^2} \partial^2\phi_3 
+ \frac{1}{v_\Delta} \frac{4v_\Delta^2 - v_\Phi^2}{v_\Phi^2 + 4
 v_\Delta^2} \partial^2\eta_3
\right)
  \varphi_{(+)} 
 \varphi_{(-)} \,
 \nonumber\\
&~
+\frac{1}{2}
\left(  2\,\frac{v_\Phi}{v_\Phi^2 + 4 v_\Delta^2} \phi_3 
+ \frac{1}{v_\Delta} \frac{4v_\Delta^2 - v_\Phi^2}{v_\Phi^2 + 4
 v_\Delta^2} \eta_3
 \right)
 \left(
 \left(\partial^2\varphi_{(+)}\right)
 \varphi_{(-)}
 +\varphi_{(+)}
 \left(\partial^2\varphi_{(-)}\right)
 \right),
\end{align}
where we ignored the total derivative.
The first term vanishes for $t\to 0$ limit.
By replacing $\partial^2\varphi_{(\pm)}$ to
$m_\varphi^2\varphi_{(\pm)}$, the effective DM-scalar interaction for
the DM-fermion scattering becomes for $t\to 0$
\begin{align}
\left(  2\,\frac{v_\Phi}{v_\Phi^2 + 4 v_\Delta^2} \phi_3 
+ \frac{1}{v_\Delta} \frac{4v_\Delta^2 - v_\Phi^2}{v_\Phi^2 + 4
 v_\Delta^2} \eta_3
 \right)
 m_\varphi^2
 \varphi_{(+)}
 \varphi_{(-)}
 &=:\left(
\begin{array}{ccc}
 \kappa_{\varphi\varphi h}&
 \kappa_{\varphi\varphi \phi_3}&
 \kappa_{\varphi\varphi \eta_3}\\
\end{array}
 \right)
 \left(
 \begin{array}{c}
  h\\
  \phi_3\\
  \eta_3\\
 \end{array}
 \right)\varphi_{(+)}\varphi_{(-)},
\end{align}
where
\begin{align}
 \left(
 \begin{array}{c}
  \kappa_{\varphi\varphi h}\\
  \kappa_{\varphi\varphi \phi_3}\\
  \kappa_{\varphi\varphi \eta_3}\\
 \end{array}
 \right) 
 =
 \frac{m_\varphi^2}{v_\Phi^2+4v_\Delta^2}
 \left(
 \begin{array}{c}
  0\\
  2{v_\Phi}\\
 \frac{4v_\Delta^2 - v_\Phi^2}{v_\Delta}\\
 \end{array}
 \right).
\label{Eq:Scalar-interaction} 
\end{align}

Next, Yukawa interaction terms of the scalar fields and SM fermions
$f$ on the vacuum are given by
\begin{align}
 \frac{m_f}{v}h\bar{f}f
 =\frac{m_f}{v}\left(
 \begin{array}{ccc}
 1&0&0 \\
 \end{array}
 \right)
 \left(
 \begin{array}{c}
  h\\
  \phi_3\\
  \eta_3\\
 \end{array}
 \right)\bar{f}f,
\label{Eq:Yukawa-interaction} 
\end{align}
where $m_f$ stands for a mass parameter of the SM fermion $f$.

By using the DM-scalar and scalar-fermion interactions in
Eqs.~(\ref{Eq:Scalar-interaction}) and (\ref{Eq:Yukawa-interaction}),
we find that the scattering amplitude
shown in Figure~\ref{Figure:Scattering}
for $t\to 0$ is given by
\begin{align}
{\cal A}_{dd}&\underset{\sim}{\propto}
  \left(
 \begin{array}{ccc}
  \kappa_{\varphi\varphi h}&
  \kappa_{\varphi\varphi \phi_3}&
  \kappa_{\varphi\varphi \eta_3}
 \end{array}
 \right)
\left(M_H^2\right)^{-1}
 \frac{m_f}{v}
 \left(
 \begin{array}{c}
  1\\
  0\\
  0\\
 \end{array}
 \right),
 \label{Eq:Add-Cross-Section-kappa} 
\end{align}
where
$\left(M_H^2\right)^{-1}
=U_H^{-1}\left[\left(M_H^2\right)^{\rm diag}\right]^{-1}U_H$;
the approximate form of $U_H$ for $v_\Delta\gg v_\Phi,v$ is given
in Eq.~(\ref{Eq:UH-approximate}).
Substituting Eq.~(\ref{Eq:Scalar-interaction}) into
Eq.~(\ref{Eq:Add-Cross-Section-kappa}), we get
\begin{align}
{\cal A}_{dd}&\underset{\sim}{\propto}
\frac{m_\varphi^2}{4v_\Delta^2}
\frac{m_f}{v}
\Bigg[
v_\Phi\sin2\alpha_z
\left(
\frac{1}{m_{h_1}^2}-\frac{1}{m_{h_2}^2}
\right)
+
4v_\Delta
\left(
-\frac{1}{m_{h_1}^2}\alpha_y \cos \alpha_z
+\frac{1}{m_{h_2}^2}\alpha_x \sin \alpha_z
\right)
\Bigg] 
 ,
\end{align}
which is identical to Eq.~(\ref{Eq:Add-Cross-Section}).

\bibliographystyle{utphys}
\bibliography{../../arxiv/reference}

\end{document}